\documentclass[twocolumn]{aastex701}

\usepackage{xspace}
\usepackage{amsmath}
\usepackage{textgreek}

\newcommand{\clname}{MACS\,J1931.8-2635\xspace}
\newcommand{\snname}{SN Eos\xspace}
\newcommand{\rmsfinal}{$0.44''$\xspace}
\newcommand{\Msol}{\ensuremath{M_{\odot}}}
\newcommand{\LCDM}{\textLambda CDM\xspace}
\newcommand{\softname}{\texttt{AstroLensPy}\xspace}

\shorttitle{VENUS: Strong-lensing model of MACS\,J1931.8-2635}
\shortauthors{J. Allingham et al.}

\tabletypesize{\footnotesize}
\begin{document}

\title{VENUS: Strong-lensing model of MACS\,J1931.8-2635 -- revealing the farthest multiply imaged supernova}


\correspondingauthor{Joseph Allingham}
\email{allingha@post.bgu.ac.il}

\author[0000-0003-2718-8640]{Joseph F. V. Allingham}
\affiliation{Department of Physics, Ben-Gurion University of the Negev, P.O. Box 653, Be'er-Sheva 84105, Israel}
\email{allingha@post.bgu.ac.il}

\author[0000-0002-0350-4488]{Adi Zitrin}
\affiliation{Department of Physics, Ben-Gurion University of the Negev, P.O. Box 653, Be'er-Sheva 84105, Israel}
\email{zitrin@bgu.ac.il}

\author[0000-0002-5588-9156]{Vasily Kokorev}
\affiliation{Cosmic Frontier Center, The University of Texas at Austin, Austin, TX 78712, USA}
\affiliation{Department of Astronomy, The University of Texas at Austin, 2515 Speedway Blvd Stop C1400, Austin, TX 78712, USA}
\email{vkokorev@utexas.edu}

\author[orcid=0009-0006-6763-4245]{Hiroto Yanagisawa}
\affiliation{Institute for Cosmic Ray Research, The University of Tokyo, 5-1-5 Kashiwanoha, Kashiwa, Chiba 277-8582, Japan}
\affiliation{Department of Physics, Graduate School of Science, The University of Tokyo, 7-3-1 Hongo, Bunkyo, Tokyo 113-0033, Japan}
\email{yana@icrr.u-tokyo.ac.jp}

\author[0000-0001-9065-3926]{Jose M. Diego}
\affiliation{Instituto e F\'isica de Cantabria,(CSIC-UC), Avda. Los Castros s/n. 39005, Santander, Spain}
\email{jdiego@ifca.unican.es}

\author[0000-0001-6278-032X]{Lukas J. Furtak}
\affiliation{Cosmic Frontier Center, The University of Texas at Austin, Austin, TX 78712, USA} 
\affiliation{Department of Astronomy, The University of Texas at Austin, 2515 Speedway Blvd Stop C1400, Austin, TX 78712, USA}
\email{furtak@utexas.edu}

\author[orcid=0000-0003-3983-5438,sname=Asada,gname=Yoshihisa]{Yoshihisa Asada}
\affiliation{David A. Dunlap Department of Astronomy and Astrophysics, University of Toronto, 50 St. George Street, Toronto, Ontario, M5S 3H4, Canada}
\email{yoshi.asada@utoronto.ca}

\author[0000-0001-7410-7669]{Dan Coe}
\affiliation{Space Telescope Science Institute, 3700 San Martin Drive, Baltimore, MD 21218, USA}
\affiliation{Association of Universities for Research in Astronomy (AURA), Inc.~for the European Space Agency (ESA)}
\affiliation{Center for Astrophysical Sciences, Department of Physics and Astronomy, The Johns Hopkins University, 3400 N Charles St. Baltimore, MD 21218, USA}
\email{dcoe@stsci.edu}

\author[0000-0003-4263-2228]{David A. Coulter}
\affiliation{William H. Miller III Department of Physics and Astronomy, Johns Hopkins University, 3400 North Charles Street, Baltimore, MD 21218, USA}
\affiliation{Space Telescope Science Institute, 3700 San Martin Drive, Baltimore, MD 21218, USA}
\email{dcoulter@stsci.edu}

\author[0000-0001-7201-5066]{Seiji Fujimoto}
\affiliation{David A. Dunlap Department of Astronomy and Astrophysics, University of Toronto, 50 St. George Street, Toronto, Ontario, M5S 3H4, Canada}
\affiliation{Dunlap Institute for Astronomy and Astrophysics, 50 St. George Street, Toronto, Ontario, M5S 3H4, Canada}
\email{seiji.fujimoto@utoronto.ca}

\author[0000-0003-2037-4619]{Conor Larison}
\affiliation{Space Telescope Science Institute, 3700 San Martin Drive, Baltimore, MD 21218, USA}
\email{clarison@stsci.edu}

\author[0000-0003-3484-399X]{Masamune Oguri}
\affiliation{Center for Frontier Science, Chiba University, 1-33 Yayoi-cho, Inage-ku, Chiba 263-8522, Japan}
\affiliation{Department of Physics, Graduate School of Science, Chiba University, 1-33 Yayoi-Cho, Inage-Ku, Chiba 263-8522, Japan}
\email{masamune.oguri@chiba-u.jp}

\author[0000-0002-2361-7201]{Justin D. R. Pierel}
\affiliation{Space Telescope Science Institute, 3700 San Martin Drive, Baltimore, MD 21218, USA}
\email{jpierel@stsci.edu}

\author[0000-0002-4622-6617]{Fengwu Sun}
\affiliation{Center for Astrophysics $|$ Harvard \& Smithsonian, 60 Garden St., Cambridge, MA 02138, USA}
\email{fengwu.sun@cfa.harvard.edu}

\author[0000-0001-5984-0395]{Maru\v{s}a Brada{\v c}}
\affiliation{University of Ljubljana, Faculty of Mathematics and Physics, Jadranska ulica 19, SI-1000 Ljubljana, Slovenia}
\affiliation{Department of Physics and Astronomy, University of California Davis, 1 Shields Avenue, Davis, CA 95616, USA}
\email{marusa.bradac@fmf.uni-lj.si}

\author[0000-0001-8460-1564]{Pratika Dayal}
\affiliation{Canadian Institute for Theoretical Astrophysics, 60 St George St, University of Toronto, Toronto, ON M5S 3H8, Canada}
\affiliation{David A. Dunlap Department of Astronomy and Astrophysics, University of Toronto, 50 St. George Street, Toronto, Ontario, M5S 3H4, Canada}
\affiliation{Department of Physics, 60 St George St, University of Toronto, Toronto, ON M5S 3H8, Canada}
\email{pratika.dayal@utoronto.ca}

\author[0000-0003-2540-7424]{Paulo A. A. Lopes}
\affiliation{Observatório do Valongo, Universidade Federal do Rio de Janeiro, Ladeira do Pedro Antônio 43, Rio de Janeiro, RJ, 20080-090, Brazil}
\email{plopes@ov.ufrj.br}

\author[0000-0002-7876-4321]{Ashish K. Meena}
\email{ashishmeena766@gmail.com}
\affiliation{Department of Physics, Indian Institute of Science, Bengaluru 560012, India}

\author[0000-0002-2282-8795]{Massimo Pascale}
\affiliation{Department of Physics \& Astronomy, University of California, Los Angeles, 430 Portola Plaza, Los Angeles, CA 90095, USA}
\email{mpascale@astro.ucla.edu}

\author[0000-0003-3596-8794]{Hollis B. Akins}
\affiliation{Department of Astronomy, The University of Texas at Austin, 2515 Speedway Blvd Stop C1400, Austin, TX 78712, USA}
\email{hollis.akins@gmail.com}

\author[0000-0002-8686-8737]{Franz E. Bauer}
\affiliation{Instituto de Alta Investigaci{\'{o}}n, Universidad de Tarapac{\'{a}}, Casilla 7D, Arica, 1010000, Chile}
\email{franz.e.bauer@gmail.com}

\author[0000-0002-7908-9284]{Larry D. Bradley}
\affiliation{Space Telescope Science Institute, 3700 San Martin Drive, Baltimore, MD 21218, USA}
\email{lbradley@stsci.edu}

\author[0000-0003-2680-005X]{Gabriel Brammer}
\affiliation{Cosmic Dawn Center (DAWN), Jagtvej 128, DK2200 Copenhagen N, Denmark}
\affiliation{Niels Bohr Institute, University of Copenhagen, Jagtvej 128, 2200 Copenhagen N, Denmark}
\email{gabriel.brammer@nbi.ku.dk}

\author[0000-0002-0302-2577]{John Chisholm}
\affiliation{Department of Astronomy, The University of Texas at Austin, 2515 Speedway Blvd Stop C1400, Austin, TX 78712, USA}
\affiliation{Cosmic Frontier Center, The University of Texas at Austin, Austin, TX 78712, USA}
\email{chisholm@austin.utexas.edu}

\author[0000-0001-8325-1742]{Guillaume Desprez}
\affiliation{Kapteyn Astronomical Institute, University of Groningen, P.O. Box 800, 9700AV Groningen, The Netherlands}
\email{g.p.a.desprez@rug.nl}

\author[0000-0001-7232-5355]{Qinyue Fei}
\affiliation{David A. Dunlap Department of Astronomy and Astrophysics, University of Toronto, 50 St. George Street, Toronto, Ontario, M5S 3H4, Canada}
\email{qyfei.astro@gmail.com}

\author[0000-0001-7113-2738]{Henry C. Ferguson} 
\affiliation{Space Telescope Science Institute, 3700 San Martin Drive, Baltimore, MD 21218, USA}
\email{ferguson@stsci.edu}

\author[0000-0001-8519-1130]{Steven L. Finkelstein}
\affiliation{Department of Astronomy, The University of Texas at Austin, 2515 Speedway Blvd Stop C1400, Austin, TX 78712, USA}
\affiliation{Cosmic Frontier Center, The University of Texas at Austin, Austin, TX 78712, USA}
\email{stevenf@astro.as.utexas.edu}

\author[0000-0003-1625-8009]{Brenda Frye}
\affiliation{Department of Astronomy/Steward Observatory, University of Arizona, 933 N. Cherry Ave, Tucson, AZ 85716}
\email{bfrye@arizona.edu}

\author[0000-0001-9411-3484]{Miriam Golubchik}
\affiliation{Department of Physics, Ben-Gurion University of the Negev, P.O. Box 653, Be'er-Sheva 84105, Israel}
\email{golubmir@post.bgu.ac.il}

\author[orcid=0000-0000-0000-0001,sname='Inayoshi']{Kohei Inayoshi}
\affiliation{Kavli Institute for Astronomy and Astrophysics, Peking University, Beijing 100871, China}
\email{inayoshi@pku.edu.cn}

\author[0000-0002-6090-2853]{Yolanda Jim\'enez-Teja}
\affiliation{Instituto de Astrof\'isica de Andaluc\'ia--CSIC, Glorieta de la Astronom\'ia s/n, E--18008 Granada, Spain}
\affiliation{Observat\'orio Nacional, Rua General Jos\'e Cristino, 77 - Bairro Imperial de S\~ao Crist\'ov\~ao, Rio de Janeiro, 20921-400, Brazil}
\email{yojite@iaa.es}

\author[0000-0002-6610-2048]{Anton M. Koekemoer}
\affiliation{Space Telescope Science Institute, 3700 San Martin Drive, Baltimore, MD 21218, USA}
\email{koekemoer@stsci.edu}

\author[0000-0003-1581-7825]{Ray A. Lucas}
\affiliation{Space Telescope Science Institute, 3700 San Martin Drive, Baltimore, MD 21218, USA}
\email{lucas@stsci.edu}

\author[0000-0002-4872-2294]{Georgios E. Magdis}
\affiliation{Cosmic Dawn Center (DAWN), Jagtvej 128, DK2200 Copenhagen N, Denmark}
\affiliation{DTU-Space, Technical University of Denmark, Elektrovej 327, 2800, Kgs. Lyngby, Denmark}
\email{geoma@space.dtu.dk}

\author[orcid=0000-0003-3243-9969]{Nicholas S. Martis}
\affiliation{University of Ljubljana, Faculty of Mathematics and Physics, Jadranska ulica 19, SI-1000 Ljubljana, Slovenia}
\email{nicholas.martis@fmf.uni-lj.si}

\author[0000-0002-9651-5716]{Richard Pan}
\affiliation{Department of Physics and Astronomy, Tufts University, 574 Boston Avenue, Suite 304, Medford, MA 02155, USA}
\email{richard.pan@tufts.edu}

\author[0000-0001-5492-1049]{Johan Richard}
\affiliation{Univ Lyon, Univ Lyon1, ENS de Lyon, CNRS, Centre de Recherche Astrophysique de Lyon UMR5574, 69230 Saint-Genis-Laval, France}
\email{johan.richard@univ-lyon1.fr}

\author[0000-0003-4223-7324]{Massimo Ricotti}
\affiliation{Department of Astronomy, University of Maryland, College Park, 20742, USA}
\email{ricotti@umd.edu}

\author[orcid=0009-0009-4388-898X]{Gregor Rihtar\v{s}i\v{c}}
\affiliation{University of Ljubljana, Faculty of Mathematics and Physics, Jadranska ulica 19, SI-1000 Ljubljana, Slovenia}
\email{gregor.rihtarsic@fmf.uni-lj.si}

\author[0000-0002-6265-2675]{Luke Robbins}
\affiliation{Department of Physics and Astronomy, Tufts University, 574 Boston Avenue, Suite 304, Medford, MA 02155, USA}
\email{andrew.robbins@tufts.edu}

\author[0000-0003-1889-0227]{William Sheu}
\affiliation{Department of Physics \& Astronomy, University of California, Los Angeles, 430 Portola Plaza, Los Angeles, CA 90095, USA}
\email{wsheu@astro.ucla.edu}

\author[0000-0003-1815-0114]{Brian Welch}
\affiliation{International Space Science Institute, Hallerstrasse 6, 3012 Bern, Switzerland}
\email{bwelch.astro@gmail.com}

\author[0000-0002-4201-7367]{Chris Willott}
\affiliation{NRC Herzberg, 5071 West Saanich Rd, Victoria, BC V9E 2E7, Canada}
\email{chris.willott@nrc.ca}

\author[0000-0001-8156-6281]{Rogier A. Windhorst}
\affiliation{School of Earth and Space Exploration, Arizona State University, Tempe, AZ 85287-6004, USA} 
\email{Rogier.Windhorst@asu.edu}


\begin{abstract}
We present a parametric strong-lensing model for the galaxy cluster \clname ($z_l = 0.35$), accompanying the detection of the spectroscopically confirmed \snname at $z = 5.13$ \citep{Coulter2026_SN_Eos}. 
We identify 10 new multiple-image systems in recent VENUS JWST/NIRCam imaging, so that the model is constrained with a total of 19 robust multiple-image systems -- nine of which also have a spectroscopic redshift. 
For the point-like source corresponding to \snname, our model predicts a total of five images, with the observed radial image pair having a similar magnification of $\mu \simeq 25 - 30$ and a small time delay of $< 5$\,days, in agreement with their simultaneous observation. According to the model, the other three predicted images arrived earlier, with time delays of $3.6 \pm 0.7$, $3.4 \pm 0.7$ and $53.9 \pm 10.8$ years prior to the two observed images, and with magnifications of $14.5 \pm 2.9$, $11.9 \pm 2.4$ and $2.2 \pm 0.4$, respectively.
The absence of detections at the predicted positions, where the host galaxy's images are also visible, confirms the transient nature of the source. 
\snname and its host galaxy are studied in separate articles, and we here focus on the lens model. The final model reaches a very good $r.m.s.$ distance between model and observations of \rmsfinal.
We present the lens-modeling results, including newly identified systems such as a triply imaged, grand-design spiral galaxy candidate at $z \simeq 3.65_{-0.09}^{+0.04}$, and discuss the potential of using high-redshift lensed SNe for cosmography.
\end{abstract}



\section{Introduction}\label{sec:intro}

Galaxy clusters are excellent laboratories for probing large-scale physics. As the most massive and expansive gravitationally tied structures of the Universe, they witness the formation of both the baryonic structures (the galaxies, the intra-cluster medium; ICM) and the elusive dark-matter scaffold \citep{Voit2005,Kravtsov_Borgani2012,Bower2013}. Massive galaxy clusters, provided the appropriate geometry, \textit{strongly lens} background sources, imaging them multiple times and introducing an arrival time delay between the multiple images. They moreover magnify these sources, sometimes by factors $\gtrsim 1000$ (in the case of compact sources near caustics, for example), thus unveiling distant, faint objects otherwise indiscernible with our best telescopes.
Owing to their usefulness for studying both their intricate structures and their magnified backgrounds, strong-lensing clusters have played important parts in modern astrophysics, such as in the study of cosmology \citep{Refsdal1964MNRAS,Jullo2010,Bonvin2017_Cosmograil,Caminha2022}, structure formation \citep{Meneghetti2013, Natarajan2017Substructure} and dark matter \citep{Clowe2006Bullet, Randall2008, Diego2018, Vegetti2024review}, but also of lensed backgrounds sources, including intriguing or high-redshift galaxies \citep{Coe2012highz,Salmon2020HighzRelics,RobertsBorsani2023highz}, stars \citep{Kelly2018NatAsCCE, Welch2022_lensedstar,Meena2023_lensedstars}, supernovae \citep{Kelly2015Sci,Rodney2016_SN_Refsdal,Frye2024SN_H0pe,Pierel2024SN_Encore,Suyu2025_SN_Encore,Pascale2025_H0pe}, active galactic nuclei \citep[AGN,][]{Furtak2023AGN,Napier2023ApJ...954L..38N,Allingham2025_SPTCLJ0546,Cloonan2025_AGNcllensed,Fei2025IMBH}, or little red dots \citep[LRD, e.g.][]{Furtak2023_LRD, Golubchik2025_LRD_VENUS, Zhang2025LRD}.
In all of these instances, the quality of the lens model is key to precisely understanding the intrinsic properties -- such as absolute magnitude, geometry, position and possible time evolution -- of the background sources.

In September 2025, JWST observed the galaxy cluster \clname with 10 NIRCam filters as part of the VENUS survey (Vast Exploration for Nascent, Unexplored Sources, \href{https://www.stsci.edu/jwst-program-info/download/jwst/pdf/6882/}{PID: 6882-GO}, PI: Fujimoto; see also section \ref{sec:data_imaging} below). The remarkable imaging quality of JWST once again proved beneficial, revealing two red, point-like images in the data. 
Based on pre-existing lens models coupled with spectroscopic data \citep[][]{Zitrin2015,Caminha2019}, a new lens model was assembled, showing the two red detections to be images of the same background source. The lens model predicted a total of \textit{five} images all within the JWST field-of-view, in contrast with the two observed.
Owing to the short time delay indicated by the model between the two detected images, and the much larger predicted delay with the unobserved predictions despite sufficient magnification, we hypothesized this source to be a transient.
The absence of a source in ancillary Hubble Space Telescope (HST) data from 2012, coupled with a detection in other HST data taken in 2024, further strengthened this scenario. Provided the exceptional combination of HST and JWST filters available, respectively from CLASH \citep{PostmanCLASH2012} and VENUS, a detailed 17-filter photometric study of the transient candidate could be undertaken, suggesting it is a supernova (SN) type II-P at an implied redshift of $\sim5$.
Inspection of archival VLT/MUSE observation revealed a $\lambda = 754.4$\,nm Ly-\textalpha\ emission line, not only around the position of the SN candidate, but also, with a $> 5\sigma$ significance, near the model-predicted position of the two additional images that lie within the MUSE field of view. This spectroscopic detection of Ly-\textalpha\ from the host galaxy at a redshift $z = 5.133 \pm 0.001$ aligned well with the high-redshift supernova hypothesis.
Indeed, a DDT proposal (\href{https://www.stsci.edu/jwst-program-info/download/jwst/pdf/9493/}{PID: 9493-DD}, PI: Coulter), adding NIRCam exposure and critically, JWST/NIRSpec spectroscopy, was undertaken in October 2025 to capture the spectrum of the transient confirming it is a distant $z=5.13$ supernova -- \textit{SN Eos}. The discovery of the supernova -- the farthest multiply lensed to date and the farthest spectroscopically observed (cf. also \citealt{Levan2025A&A...704L...8L}) -- is presented in our companion article \cite{Coulter2026_SN_Eos}.

In the present article, we focus on the lens model that accompanied the detection of \snname (SN 2024aijd), taking advantage of the combined VENUS and CLASH datasets, as well as data available from VLT/MUSE. These observations, their photometry, and past analyses are described in \S \ref{sec:data}. The lensing analysis of the cluster, including the underlying method, software, choices, constraints, and unique multiple images identified, is described in \S \ref{sec:analysis}. 
We discuss the lens-modeling results in \S \ref{sec:discussion}, and dedicate simulations based on our realistic lens-modeling to inspect how distant supernovae may be useful to constrain cosmology.
The article is then summarized in \S \ref{sec:summary}.

Throughout this paper, unless stated otherwise, we use a standard flat \LCDM model with $H_0=70\,\mathrm{km}\,\mathrm{s}^{-1}\,\mathrm{Mpc}^{-1}$, $\Omega_m=0.3$ and $\Omega_{\Lambda}=0.7$ for cosmology, the AB system \citep{Oke1983ABandStandards} for magnitudes and $1\sigma$ ranges for uncertainties. With the above cosmological parameters, one arcsecond implies a scale of 4.96\,kpc at the cluster's redshift ($z_l = 0.35$) and of $6.20$\,kpc at the redshift of \snname ($z_s = 5.13$).

\section{Data}\label{sec:data}

The galaxy cluster \clname (also known as ACT-CL J1931.8-2634) was detected in a ROSAT X-ray selection \citep{EbelingMacsCat2001}. Its remarkable lensing features, identified in the framework of the CLASH survey \citep[][see below]{PostmanCLASH2012,Zitrin2015}, rendered it a valuable target for the VENUS survey of strong-lensing clusters with JWST. We describe here the data used in this work, as well as some other key data available for the cluster, highlighting some of its properties and potentially aiding future analyses of the cluster.

\subsection{X-rays, radio and sub-millimeter ancillary observations} \label{sec:Xray_radio_submm}

\clname was imaged with the \textit{Chandra} X-ray telescope \citep[see][]{Ehlert2011MACS1931}, revealing one of the most X-ray luminous cool-cores. Its relaxed ICM morphology was further characterized as part of the CLASH survey \citep{Donahue2016}, using a combination of X-ray and Sunyaev-Zel'dovich effect (SZE) sub-millimeter observations.
In the most recent Atacama Cosmology Telescope SZE observation \citep[ACT DR6,][]{ACT_DR6}, the cluster's mass was constrained to $M_{200,c} = 1.09_{-0.18}^{+0.21} \times 10^{15} \Msol$. 
While these studies probe the thermodynamic and dynamical state of the cluster, they provide a valuable multi-wavelength, larger-radius context for our strong-lensing analysis, which instead traces the projected mass distribution in the cluster core.

The cluster has also been the subject of detailed combined radio \citep{Ehlert2011MACS1931, 2018ApJ...853..100Y, 2021MNRAS.504.2924T} and X-ray observations of the AGN hosted by its brightest cluster galaxy (BCG). These investigations suggest a north-south oscillation of the cool-core ICM around the BCG, conversely to the east-west AGN emission jets, estimated to be amongst the most powerful known. 
In another example, \cite{2019ApJ...879..103F} used Atacama Large Millimeter Array (ALMA) observations of \clname, finding the BCG to be an exceptionally massive cold gas reservoir, including very cold dust clumps ($< 10$\,K).

\begin{figure*}
\includegraphics[width=1\textwidth]{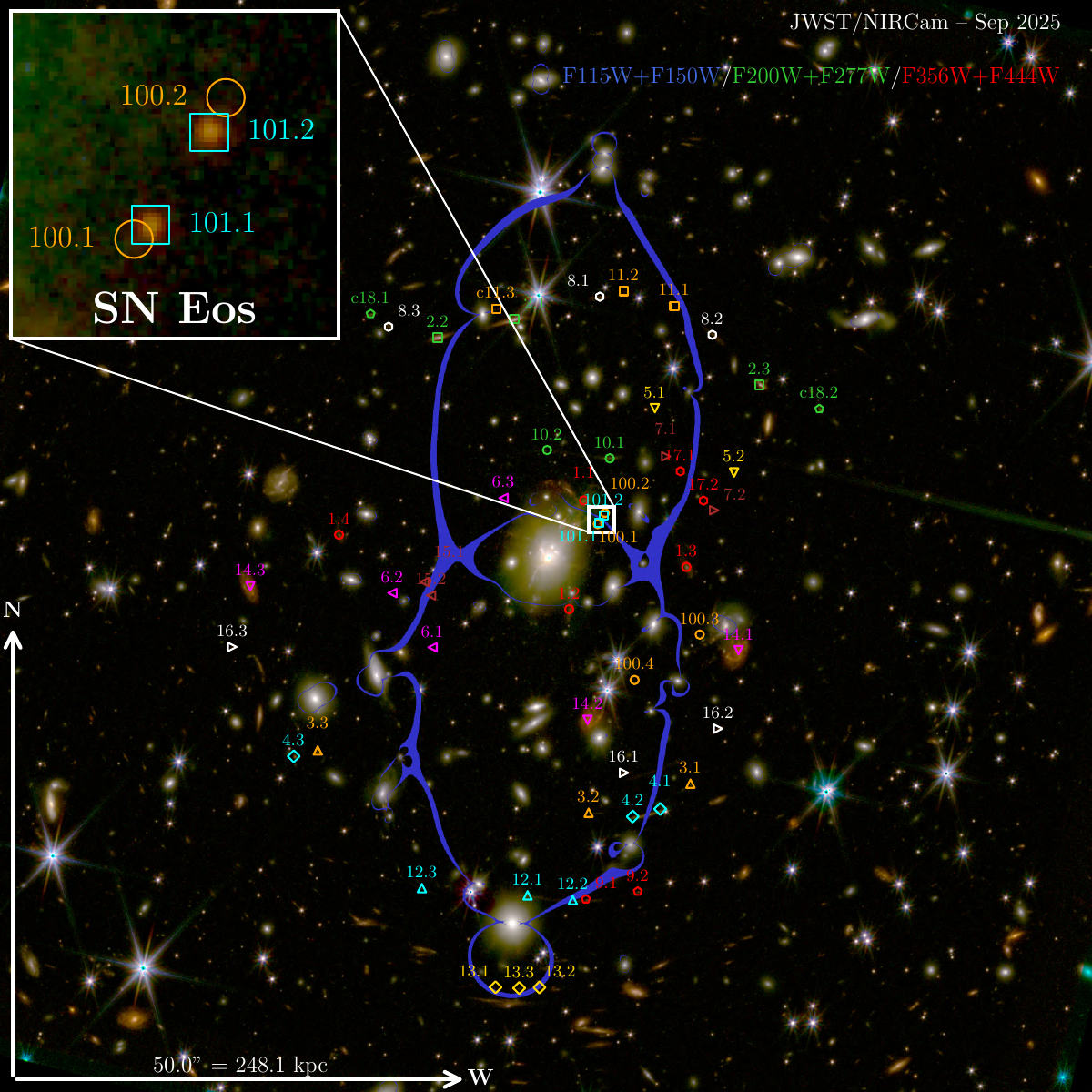} 
\caption{Composite color image of \clname, using the summed VENUS-collected JWST NIRCam bands F115W and F150W for the blue channel, F200W and F277W for the green, and F356W and F444W for the red channel.
All 20 strongly-lensed systems are represented with a unique shape-color couple. This includes candidate system 18, although it is not used as a constraint. We show the Eos galaxy host and the SN (respectively, systems 100 and 101) in a zoomed-in image on the top-left corner.
The blue lines represent the $|\mu| > 100$ critical lines for \snname, i.e.\ a source redshift of $z_s = 5.13$. In the lens plane, $50''$ represent $248.1$ physical kpc.}
\label{fig:Colour_image_m1931}
\end{figure*}

\subsection{Optical and NIR imaging data} \label{sec:data_imaging}

The continued interest in \clname, together with its strong-lensing features \citep{Zitrin2015,Caminha2019}, led to its inclusion in the VENUS survey \citep{fujimoto_venus_2025}. The 10 NIRCam bands (F090W, F115W, F150W, F200W, and F210M in the short-wavelength channel, and F277W, F300M, F356W, F410M, and F444W in the long-wavelength channel) collected in 2025 September amount to a total exposure time of 4.6\,hours. The observations achieve $5\sigma$ depths of 28 AB magnitudes per band. \href{https://www.stsci.edu/jwst-program-info/download/jwst/pdf/9493/}{DDT proposal 9493} (PI: Coulter) on \snname added four-band, multi-epoch NIRCam observations (amounting to 2319\,s of observation in F070W, and 773\,s in the F277W, F356W and F444W bands) taken in 2025 October. 

The data have been reduced as follows. We begin with JWST level-2 products from MAST, consisting of all the individual single-exposure images, already calibrated with stages 1 and 2 of the JWST Pipeline\footnote{\url{https://github.com/spacetelescope/jwst}} \citep{Bushouse2022JWSTPipeline}, and process them further with the \texttt{grizli} pipeline \citep{brammer_grizli_2023}, as was done for the DAWN JWST Archive \citep[DJA;][]{valentino22}\footnote{\url{https://dawn-cph.github.io/dja/}}. We perform further post-processing steps that include additional background, $1/f$ noise and diffraction spike subtraction procedures, both at the amplifier level for each filter, and then the final drizzled mosaic \citep[][]{endsley24,kokorev25a}. The final NIRCam mosaics were drizzled to the respective detector pixel scales ($0.03$\,arcsec/pix for the SW channel and $0.06$\,arcsec/pix for the LW channel).
Additionally, JWST/MIRI IFU and F2100W image data exist around the BCG (PID: 3629-GO, PI: Man), manifesting the interest in the cluster's BCG \citep{Ghodsi2026}, although we do not make use of these here.

We supplement our JWST dataset with ancillary archival HST imaging from the Cluster Lensing And Supernova survey with Hubble \citep[CLASH, \href{https://archive.stsci.edu/proposal_search.php?mission=hst&id=12456}{PID: 12456}, PI: Postman,][]{PostmanCLASH2012}, including 7 ACS bands, 5 WFC3/IR bands and 4 WFC3/UVIS bands (amounting to a total 20 orbits $\simeq 10,000$\,s exposure), and from more recent observations in the WFC3/UVIS F814W and WFC3/IR F110W filters taken in March 2024, as part of the lensed stars SNAP survey \citep[PID: 17504,][]{kelly_snap_2023}. 
The WFC3 observations are only used for a by-eye inspection and galaxy selection.
The CLASH observations were treated uniformly as part of the Complete Hubble Archive for Galaxy Evolution initiative (CHArGE), as described in \cite{kokorev22}.

Although not used here, we note \clname cluster observations with Subaru in 2006 and 2007 for a total $\gtrsim 10,000$\,s in five bands, also used as part of CLASH \citep{Umetsu2014CLASH_WL,Merten2014CLASHcM}. For example, through weak gravitational lensing jointly with magnification measurements, \cite{Umetsu2018} estimate the cluster mass to be $M_{200,c} = (1.16 \pm 0.28) \times 10^{15} \Msol$.
%

\subsection{Ground and space-based spectroscopy} \label{sec:spectroscopy}

\clname was observed in 2015 June by the VLT/MUSE IFU \href{https://archive.eso.org/wdb/wdb/eso/sched_rep_arc/query?progid=095.A-0525(A)}{GTO program 095.A-0525} \citep[PI: Kneib,][]{2014Msngr.157...13B}, in three 2924\,s exposures, centered on the BCG.
These data were already used in \cite{Caminha2019} for strong lensing analysis -- and first analyzed in \cite{Rexroth2017}. In JWST data, we detect several new multiple-image systems (see Section\,\ref{sec:lensmodeling}), but only one of these is detected spectroscopically in the MUSE data -- the host galaxy of \snname at $z=5.133$, with four images detected in the MUSE 1\,arcmin $\times$ 1\,arcmin field-of-view, listed in Table\,\ref{tab:m1931_mulim} (System 100).
From the galaxy velocity dispersion, measured as part of CLASH-VLT (\href{https://archive.eso.org/wdb/wdb/eso/sched_rep_arc/query?tel=UT3&from_date=01-Apr-2012&progid=186.A-0798(F)&period=89&remarks=Type:%20Large}{LP186.A-0798}, PI: P.\ Rosati), a mass $M_{200,c} = 7.96_{-1.00}^{+1.51} \times 10^{14} \Msol$ was measured for the cluster (P.\ A.\ Lopes et al., private communication), consistent with the aforementioned estimate based on the SZE, but lying a bit beyond $1\sigma$ from the weak lensing CLASH measurement by \cite{Umetsu2018}.
This discrepancy, while not statistically significant, may reflect systematic differences in the methods or assumptions of each technique, as each probes different physical properties of the cluster (galaxy kinematics, gas dynamics, or gravitational potential).

In addition, NIRSpec/PRISM and CLEAR spectroscopic observations were executed with JWST on 2025 October 8th (PID: 9493-DDT, PI: Coulter) for both observed images of \snname (System 101), amounting to a total 10,504 s of exposure each. The analysis of these data is detailed in \cite{Coulter2026_SN_Eos}; here we only make use of the spectroscopic redshift of \snname at $z=5.133 \pm 0.001$ derived from these, in accordance with the MUSE redshift for the host. We note that the analysis of the host is presented in Y.\ Asada et al.\ (in prep.).

\subsection{Red-sequence galaxy selection} \label{sec:red_sequence}

We make use of both the imaging and ground-based spectroscopy in order to isolate the red-sequence of cluster member galaxies \citep{GladdersYee2000Finder}. We run \texttt{SExtractor} \citep{BertinArnouts1996Sextractor} on the images and we use the MUSE-detected cluster members to guide the red-sequence selection, selecting objects with WFC3/F160W $m_{\rm F160W} < 23$ AB and a $m_{\rm F105W} - m_{\rm F160W}$ color lying within 3 $\sigma_{C}$ from the red sequence line (where $\sigma_{C}$ is the color standard deviation; i.e., scatter). We additionally make use of a red sequence defined using the ACS/F606W-F814W color in a similar way. The combined catalog is assembled with equal weights provided to both sequences, and cleaned up manually from spurious detections (for more details on the procedure see \citealt{Allingham2023}).
The resulting cluster-galaxy catalog contains 107 galaxies, where we use the WFC3/F160W magnitudes for scaling the galaxies in the modeling (see Section \ref{sec:analysis}).

\subsection{Photometric redshifts} \label{sec:photo-z}

Since the lensing of a source depends on its redshift, photometric redshifts ($z_{\rm phot}$) are key to lens modeling -- especially for systems lacking spectroscopic redshifts.
For this study, we adopt the VENUS \texttt{eazy} photometric redshift catalog, selected from a suite of $z_{\rm phot}$ catalogs released by the VENUS collaboration. The construction of this catalog is outlined hereafter.

Photometry in different bands, for each object, is obtained using \texttt{sep} \citep{sep}. To perform the initial detection, we start by constructing an inverse variance-weighted stack of the long wavelength (LW) mosaics by using the native F277W, F356W and  F444W mosaics. The photometry itself is measured on PSF homogenized images, all convolved to the resolution of the F444W band and is measured using \texttt{sep} in a range of circular apertures ($D = 0.1$–$1.2''$). As our fiducial aperture, we adopt the $D = 0.2''$ flux densities. The photometric uncertainties are derived directly from the variance maps, which take into account both random and correlated noise. Aperture corrections are derived as a ratio between the flux density within a given aperture and the total (Kron) flux. An additional correction to account for the missing flux outside of the circularized Kron radius is derived by following the methods outlined in \cite{Whitaker2011} and \cite{Weaver2024}.

In order to derive the $z_{\rm phot}$ from the \texttt{sep}-extracted photometry, we start with aperture-corrected JWST and HST/ACS photometry in $D=0.2''$ apertures and fit it with the \texttt{BLUE\_SFHZ\_13} template set using the spectral energy distribution (SED) fitting software \texttt{eazy} \citep{BammerEazy2008}.
The linear combinations of log-normal star formation histories included in the template set are not allowed to exceed redshifts that start earlier than the age of the Universe \citep[for more details, see][]{BlantonRoweis2007}. These models are further complemented by a blue galaxy template, derived from a JWST spectrum of a $z = 8.50$ galaxy with extreme emission-line equivalent widths \citep[EW; ID4590;][]{Carnall2023}.
We complete this VENUS photometric catalog by rerunning \texttt{eazy} with priors specific to the multiply-lensed images, and with a by-eye inspection of outlier bands presenting obvious artifacts, which we manually remove before running \texttt{eazy}.

The final photometric redshifts are given in Table \ref{tab:m1931_mulim}, for systems lacking spectroscopic redshifts.
In some cases, the photometric redshifts of different images of the same system do not overlap. This is in most cases a combination of the faintness of the images, and possible contamination by foreground galaxies.
In cases where the photometric redshifts were too inconsistent to supply useful information such as for Systems 11 and 13, these were removed from the Table. In all cases, the photometric redshifts listed here should be interpreted only as useful priors for lens modeling. The redshifts of these systems are left free to be determined in the modeling. These model redshifts can in turn be compared to the photometric ones.

\subsection{Spectral modeling} \label{sec:spec-modelling}

To derive the physical properties of lensed galaxies of interest, we use the VENUS catalog made with \texttt{Bagpipes} \citep{Bagpipes2018} for its SED fitting capabilities. In this catalog, based on the same photometric extraction as described in Section\,\ref{sec:photo-z}, the stellar component is modeled using Binary Population And Spectral Synthesis (BPASS) stellar population synthesis templates \citep{Stanway2018}, assuming delayed-$\tau$ star-formation histories. Uniform priors are imposed on the stellar age $t_\mathrm{age} \in [0.01,5]\,\mathrm{Gyr}$, the star-formation timescale $\tau \in [0.1,3]\,\mathrm{Gyr}$, the stellar metallicity $Z_\star \in [0.1,1]\,Z_\odot$, and the logarithmic stellar mass $\log_{10}(M_{{\star}}/\mathrm{M_\odot}) \in [5,13]$. Nebular emission is modeled using version 23 of the \texttt{Cloudy} photoionization code \citep{Cloudy90,Cloudy23.01}, where the BPASS spectra are adopted as the input ionizing source to generate grids of nebular line and continuum emission. A flat prior is assumed for the ionization parameter, $\log_{10}U \in [-4,-1]$, and the nebular metallicity is set to the stellar metallicity. Dust attenuation is treated using the extinction curve of \citet{Calzetti1994}, with a uniform prior on the visual extinction $A_\mathrm{V} \in [0, 5]\,\mathrm{mag}$. Redshifts are allowed to vary within the range $z\in[0,10]$. 
Specifically, we use here the \texttt{Bagpipes} results to examine the physical properties of the multiply-lensed, spiral galaxy Charybdis (hence $z < 10$, see Section\,\ref{sec:multiple_images}).

\section{Lensing Analysis}\label{sec:analysis}

\begingroup
\renewcommand{\arraystretch}{1.2}
\begin{table*}
	\caption{Best-fit parameters of the strong-lensing mass model. Columns are as follows: (i) Component name; (ii) and (iii) Displacement, in arcsec, in respectively R.A.\ ($\alpha$) and declination ($\delta$) from the cluster center $\boldsymbol{\theta_c} = (292.957083, -26.576100)$\,deg (J2000); (iv) Ellipticity $e = 1 - r_b/r_a$ (= flattening $f$), where $r_a$ and $r_b$ are respectively the semi-major and minor axes; (v) Position angle, $\theta$, in degrees, oriented directly from the astronomical west; (vi) Core radius, $a$, in kpc; (vii) Cut radius, $s$, in kpc; and (viii) Velocity dispersion, $\sigma$, in km\,s$^{-1}$. Errors include only the statistical uncertainties, and values without error bars were kept fixed.}
	\label{tab:best_model}
    \centering
    \hspace{-2cm}
	\begin{tabular}{lccccccc}
	    \hline
		\hline
		& $\Delta_{\alpha}$ [arcsec] & $\Delta_{\delta}$ [arcsec] & $e$ & $\theta$ [deg] & $a$ [kpc] & $s$ [kpc] & $\sigma$ [km s$^{-1}$]\\
		\hline
		\hline
        DM halo & $1.33_{-0.02}^{+0.04}$  & $-0.17_{-0.18}^{+0.05}$ & $0.551_{-0.003}^{+0.005}$ & $88.07_{-0.08}^{+0.18}$ & $75.6_{-0.2}^{+1.9}$ & $\infty$  &  $1172_{-5}^{+4}$ \\
        BCG & 0.80 & 0.90 & $0.393_{-0.008}^{+0.008}$ & 68.57 & $11.4_{-0.1}^{+0.0}$ & $167.5_{-1.7}^{+0.4}$  & $608.2_{-3.1}^{+0.7}$ \\
        $L^{\star}$ Galaxy catalog &  &  &  &  & $0.20$ & $17.1_{-0.8}^{+0.6}$ & $194.3_{-0.4}^{+6.0}$ \\ 
		\hline
		\hline
	\end{tabular}
\end{table*}
\endgroup

\subsection{Lens modeling software: \softname}

In this work, we construct a strong-lensing model for the galaxy cluster \clname using a new \texttt{Python} implementation of the parametric method by \citet[][]{Zitrin2015}. This method, sometimes dubbed \texttt{PIEMDeNFW}, was recently revised to allow for greater speed and higher resolution results, i.e.\ it is not limited by a pre-defined grid's resolution -- and since then has been often labeled \texttt{ZitrinAnalytic}. \texttt{ZitrinAnalytic} was written in \texttt{MATLAB}, and its performance was showcased on a variety of JWST clusters \citep[e.g.][see also \citealt{Furtak2023UNCOVER}, where more comprehensive details of the method are given]{pascale22, Meena2023_lensedstars, RobertsBorsani2023highz,Golubchik2025_LRD_VENUS}. We name the new \texttt{Python3} implementation \textbf{\softname}\footnote{The software is available through private communication.}. \softname mostly adopts the main principles of \texttt{ZitrinAnalytic}, following the original architecture and translating the various functions into \texttt{Python}. A fast computation is made possible by using \texttt{Numba} \citep{numba:2015, Numba_17874185}, compiling \texttt{Python} functions into optimized machine code. A notable difference in \softname is the MCMC sampler: \texttt{ZitrinAnalytic} uses a custom-made MCMC based on the Metropolis–Hastings algorithm with annealing, while \softname can also use \texttt{emcee} \citep{emcee_10996751}, as is done here, to perform a MCMC-like algorithm for exploring the parameter space through likelihood gradient-guided stochastic moves \citep[cf.][]{GoodmanWeare2010StretchMove}.

Cluster lens models typically require a combination of a cluster-scale dark matter component, and a component representing the influence of individual galaxies.
For the dark matter, \softname employs large-scale, smooth dark matter halos (DMHs), modeled each with a Pseudo-Isothermal Elliptical Matter Distribution (PIEMD, cf.\ \citealt{Keeton2001models}).
For the galaxy component, individual galaxies (obtained in Section\,\ref{sec:red_sequence}), are modeled as a collection of dual Pseudo-Isothermal Elliptical profiles \citep[dPIE; see][]{KassiolaKovner1993, Eliasdottir2007arXiv0710.5636E}. 
Provided that we scale all cluster-member galaxies using the Faber-Jackson relationship for red elliptical galaxies \citep[][see Appendix\,\ref{sec:pot_in_astrolemonpy}]{FaberJackson}, the galaxy component thus requires only two free parameters: the `scale radius' $s_{\star}$, and velocity dispersion $\sigma_{\star}$, of a reference $L^{\star}$ galaxy.
Often it is desirable to model individual galaxies independently -- for example bright central galaxies playing a significant role in the lens model. \softname thus allows to optimize the luminosity of individual galaxies independently of the Faber-Jackson relation adopted.
Moreover, the ellipticity, position angle and core radius of specific galaxies may be freed. 
Appendix\,\ref{sec:pot_in_astrolemonpy} provides more details on the computation of lensing potentials (or mass distributions) in \softname.

The model optimization is done via $\chi^2$ minimization, reducing the distance between the observed and predicted image positions. In principle, constraints such as parity, relative magnification and time delays for multiple images can also be incorporated, but these are not used here. 
In practice, the $\chi^2$ minimization is done in the source plane following \citet[][properly taking into account the magnification to resize source plane distances, in order to avoid a bias towards flatter models]{2010GReGr..42.2151K}. For the final model, all relevant quantities such as best-fit $\chi^2$ or $r.m.s.$ are computed and cited in the image plane.

\begin{figure}
    \centering
    \includegraphics[width=\linewidth]{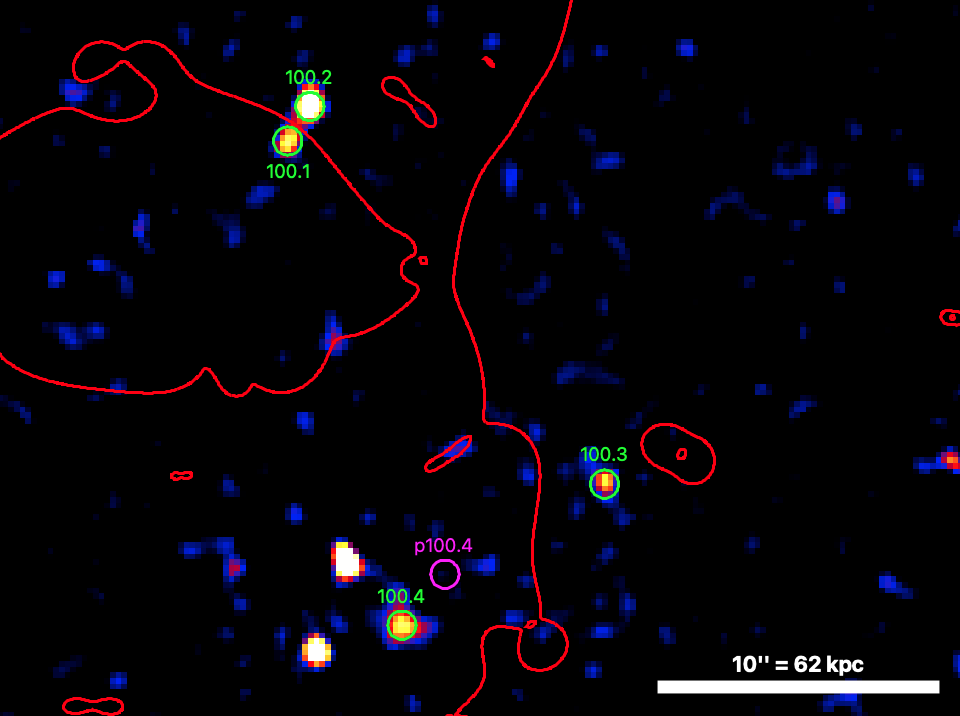}
    \caption{VLT/MUSE smoothed narrow-band at $\lambda = 7455\AA$, displaying the $z=5.133\pm 0.001$ detection of the `Eos' Ly-\textalpha\ emitting galaxy, host to \snname. In red, we display the $z=5.133$ critical lines, identical to the ones displayed in Fig.\,\ref{fig:Colour_image_m1931}. In green, we display the four $\geq 2\sigma$ detections of multiple images of galaxy Eos, while in magenta we display the predicted location for image 100.4, with a positional $\simeq 2''$ offset from the detection. Images 100.1-3 are all reproduced below pixel prediction. Note that two bright detections near 100.4 correspond to stars.}
    \label{fig:Lya_NB_Sys100}
\end{figure}

\subsection{MACS J1931.8-2635 lens modeling} \label{sec:lensmodeling}

For the galaxy cluster \clname, the galaxy component is modeled as described above, and the dark matter distribution is modeled with a single DMH, in line with the relaxed, cool-core nature of the cluster.
Given the strong gravitational role of the BCG \citep[cf.\ e.g.][]{2013ApJ...765...24N}, we leave its ellipticity as a free parameter while fixing its position angle, and free the BCG weight, so that the BCG may attain effective velocity dispersions as high as the values measured in \cite{Ciocan2021}; we thus permit the scaling parameter, which multiplies the BCG luminosity, to vary by up to a factor of 20.

\begin{figure*}
    \centering
    \includegraphics[width=0.49\textwidth]{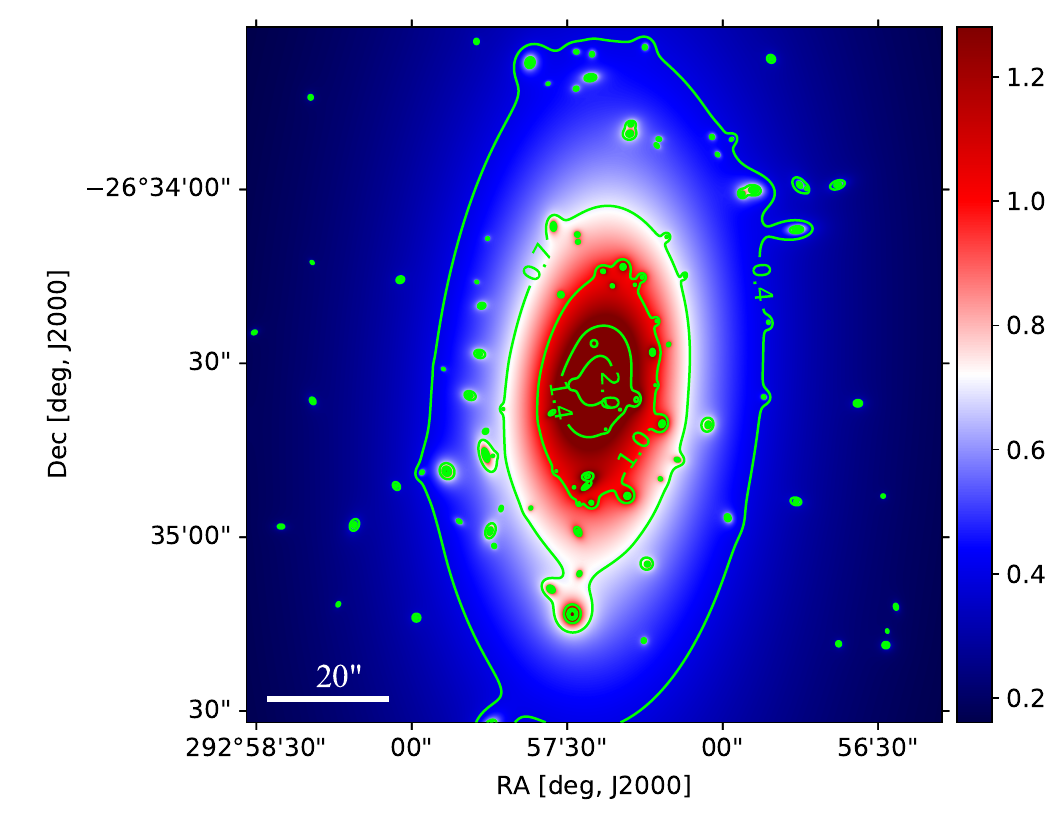}
    \includegraphics[width=0.49\textwidth]{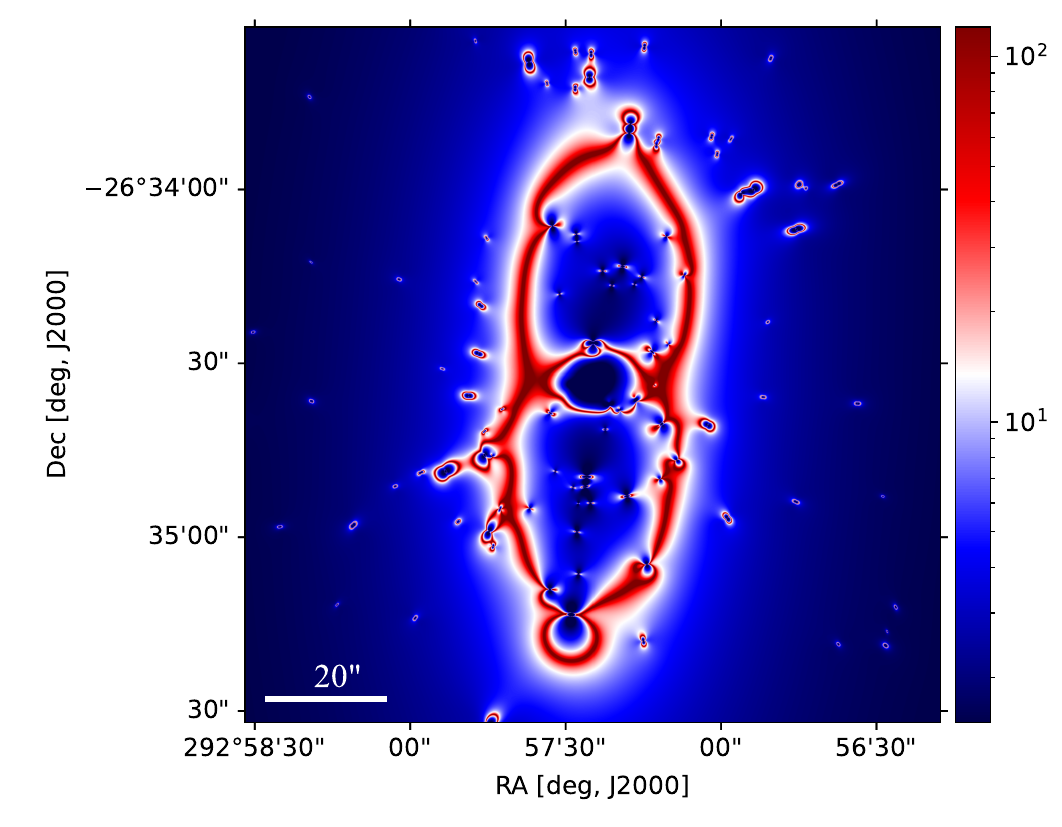}
    \caption{\textit{Left:} Convergence $\kappa$ map; \textit{Right:} Magnification $|\mu|$ map, both for the best-model at redshift $z_S = 5.13$, corresponding to the Eos system.}
    \label{fig:m1931_kappa_mu}
\end{figure*}
The model therefore includes 11 free parameters describing the mass distribution:
the velocity dispersion ($\sigma_{\star}$) and cut-radius ($s_{\star}$) of a typical L$^{\star}$ galaxy (where we adopt an absolute magnitude $M_{\star} = -22.6$), for the galaxy scaling relations; the DMH density parameters, i.e., velocity dispersion ($\sigma_{\rm DMH}$) and core radius ($a_{\rm DMH}$), and positional parameters  -- central coordinates offset ($\Delta_{\alpha, \rm DMH}$, $\Delta_{\delta, \rm DMH}$), ellipticity ($e_{\rm DMH}$) and position angle ($\phi_{\rm DMH}$); and the BCG core radius ($a_{\rm BCG}$), ellipticity ($e_{\rm BCG}$), and luminosity scaling factor ($w_{\rm BCG}$). No external shear was added.

The lens model is constrained using 19 multiply imaged systems of background sources. 
Five of these (Systems 2, 3, 4, 8 and 9 here) were identified by \citet{Zitrin2015} in the framework of the CLASH program -- where spectroscopic redshifts for systems 2 and 4 were also published. \cite{Caminha2019} later found four additional systems (1, 5, 6 and 7 here), and compiled spectroscopic redshifts for seven systems (1-7) through MUSE spectroscopic analysis (see Section\,\ref{sec:data}).
Here, we considerably update the lens model using 10 new systems. These were iteratively identified using the lens model, their color, redshift, and symmetry seen in JWST/NIRCam imaging taken for the VENUS program (see Table\,\ref{tab:m1931_mulim}). This amounts to a total of 51 lensed images, spanning 19 systems -- including 9 with spectroscopic redshifts -- all marked on Fig.\,\ref{fig:Colour_image_m1931} and used to constrain the final model. In Fig.\,\ref{fig:Colour_image_m1931} we also mark a candidate system (System c18), which is not used as a constraint for the modeling.
The redshifts of all 10 systems that lack a spectroscopic detection are left free, although we provide here their photometric redshifts (see Table\,\ref{tab:m1931_mulim}).
In total, the model comprises $21$ free parameters: 11 for the mass distribution of the cluster and 10 to account for the free redshifts.
For the $\chi^{2}$ minimization, we adopt a positional error of $\sigma_{pos}= 0.2''$ per image, for all systems, except for \snname, for which we adopt a positional uncertainty of $\sigma_{pos}= 0.02''$ per image -- giving it substantial weight in the modeling so we can more accurately predict its magnification and time delays.

\begingroup
\renewcommand{\arraystretch}{1.2}
\begin{table*}
\caption{Summary of the main properties of the lens model of \clname. Columns are: (i) Redshift of the source; (ii) Effective Einstein radius (arcsec); (iii) Projected mass contained within the area enclosed by the critical lines ($10^{14} \Msol$); (iv), (v) and (vi) Area in the field for which the magnifications are resp.\ $|\mu|\geq 3$, $|\mu|\geq 5$ and $|\mu|\geq 10$ (in arcmin$^2$).}
\label{tab:m1931_lens_properties}
\centering
\hspace{-2cm}
    \begin{tabular}{l|ccccc}
    \hline
    \hline
    $z_s$ & $\theta_E$ & $M(\in \mathcal{A})$ & $\mathcal{A}_{|\mu|\geq 3}^{\rm lens}$ & $\mathcal{A}_{|\mu|\geq 5}^{\rm lens}$ & $\mathcal{A}_{|\mu|\geq 10}^{\rm lens}$\\
    & [arcsec] & [$10^{14} \Msol$] & [arcmin$^2$] & [arcmin$^2$] & [arcmin$^2$]\\
    \hline
     2 & $22.4 \pm 2.2$ & $0.80 \pm 0.12$ & 1.36 & 0.76 & 0.36\\ 
     5 & $25.7 \pm 2.6$ & $0.96 \pm 0.14$ & 1.61 & 0.94 & 0.46\\
     9 & $26.7 \pm 2.7$ & $1.01 \pm 0.15$ & 1.68 & 1.00 & 0.50\\
    \hline
    \hline
    \end{tabular}
\end{table*}
\endgroup

\subsection{Resulting lens model} \label{sec:final_lens_model}

\begin{figure*}
    \centering
    \begin{minipage}[t]{0.562\textwidth}
    \centering
    \includegraphics[width=\linewidth]{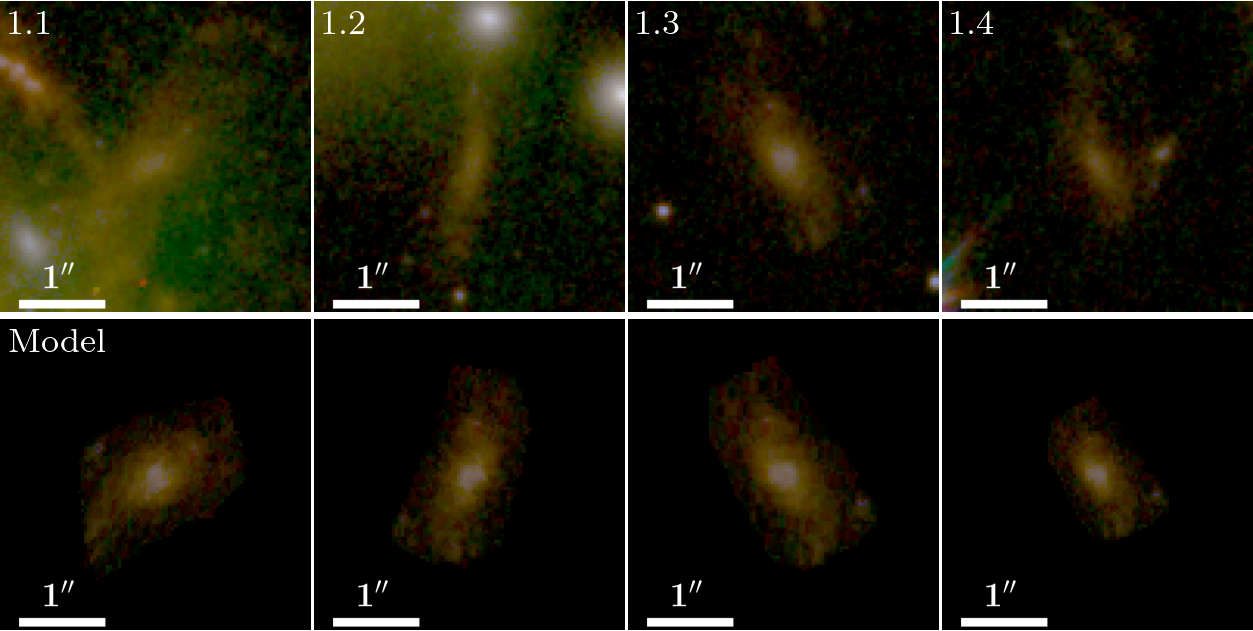}\label{subfig:Sys1}
    \end{minipage}
    \hfill
    \begin{minipage}[t]{0.422\textwidth}
    \includegraphics[width=\linewidth]{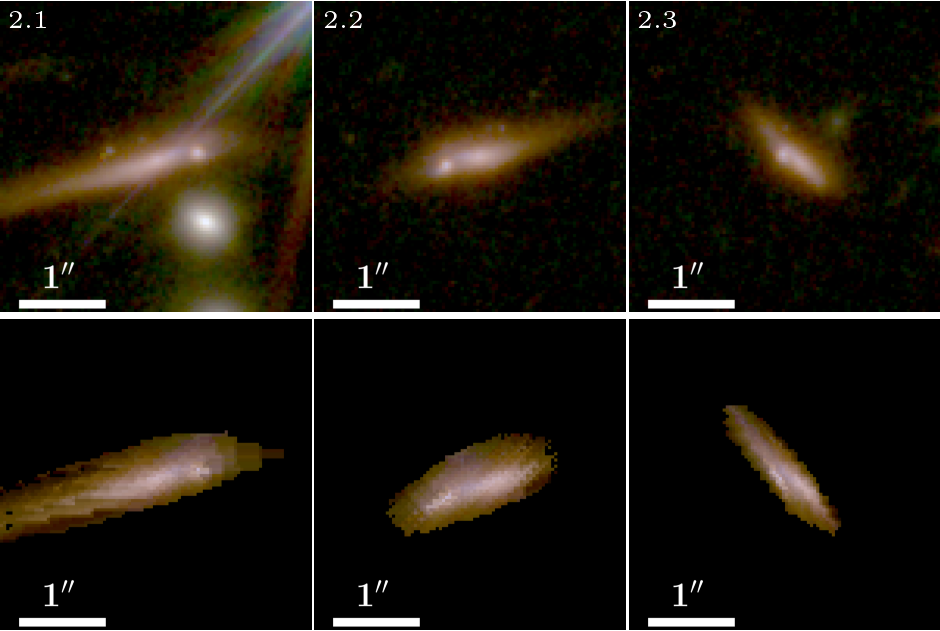}\label{subfig:Sys2}
    \end{minipage}  
    \caption{Image reconstruction of the quadruply-lensed System 1 (left) and triply-lensed System 2 (right), at respective redshifts $z_{\rm 1, spec} = 1.178$ and $z_{\rm 2, spec} = 1.835$.
    The first row presents the extracted images from Fig.\,\ref{fig:Colour_image_m1931}; the second row the image reconstruction.
    All insets are $4''\times4''$.}
    \label{fig:mulim1_mulim2}
\end{figure*}

Our final implementation converged after $< 2,000$ steps of burn-in, with 100 walkers each, within $\sim 20$\,minutes running on one CPU, using the \texttt{emcee} sampler. With 51 images over 19 systems to constrain the $21$ free parameters, we have $64$ constraints on the model\footnote{The number of strong lensing constraints is $2 (N_{\rm img} - N_{\rm src})$, where $N_{\rm img}$ is the number of multiple images, and $N_{\rm src}$ the number of multiply-lensed sources.}, resulting in $\nu = 43$ degrees-of-freedom.
With a rescaled positional uncertainty of $\sigma_{pos} = 0.5''$ per image, commonly adopted for lens modeling and thus useful for comparison, we find a best-fit model reduced $\chi_{\nu}^2 = \chi^2 / 43 \simeq 1$.
The model reproduces very well all images, with the final $r.m.s.$ error yielding \rmsfinal on average for all 19 systems, despite the large number of constraints.
In order to assess the source-plane optimization efficiency, we ran similar image-plane optimization, and found comparable convergence to the source-plane one, with best-fit parameters, magnifications, and time-delays within the $1\sigma$ errors of the one obtained in the source-plane optimization used throughout.

The model predicts a total of five images for \snname and its host galaxy, including the two images visible in VENUS data for \snname (101.1-2). 
For system 100 -- the SN host (`galaxy Eos') -- the Ly-\textalpha\ emission line is observable at 7455 \AA\ in VLT/MUSE archival data for images 100.1-.4, as displayed on Fig.\,\ref{fig:Lya_NB_Sys100}, while the fifth predicted image (p100.5) is out of the MUSE field-of-view, as is also described in Section\,\ref{sec:spectroscopy}. Moreover, the NIRCam/F300M and F410M summed observations reveal a faint photometric trace of galaxy Eos.
The existing images of Systems 100 and 101 are faithfully reproduced by our lens model within $\left|\Delta \theta \right| < 0.1''$, except for the fourth predicted image, 100.4, for which the reproduction is $\simeq2''$ away from its MUSE-observed position. Nonetheless, the two SN images, 101.1 and 101.2 are reproduced with high fidelity.

We show in Fig.\,\ref{fig:Colour_image_m1931} a RGB pseudo-color image of the cluster from VENUS data, overlaid with the $|\mu| \geq 100$ critical curves for \snname (or a source at $z_s = 5.13$). We display, for the same source redshift, the corresponding convergence and magnification maps in Fig.\,\ref{fig:m1931_kappa_mu}.
We present the optimized lens-model parameters in Table\,\ref{tab:best_model}, except for the optimized redshifts, transcribed in Table\,\ref{tab:m1931_mulim}. 
The posterior distribution for the different parameters is displayed in Fig.\,\ref{fig:cornerplot_vA_52}.

We find a final projected mass within 200\,kpc from the BCG $M_{\rm 2D}(R<200\,\mathrm{kpc}) = (1.64 \pm 0.25) \times 10^{14}\,\Msol$. 
In Table\,\ref{tab:m1931_lens_properties}, we summarize some of the lens model properties, such as the effective Einstein radius\footnote{Defined as $\theta_E = \sqrt{\mathcal{A}/\pi}$, where $\mathcal{A}$ is the critical area, i.e.\ the area enclosed within the tangential critical lines.} or the enclosed projected mass. These are later discussed in Section \ref{sec:discussion}. We also make the lens model publicly available\footnote{\dataset[doi: 10.5281/zenodo.18646525]{https://doi.org/10.5281/zenodo.18646525}}.

\subsection{Reproduction of multiple images} \label{sec:multiple_images}

\begin{figure}
    \centering
    \includegraphics[width=\linewidth]{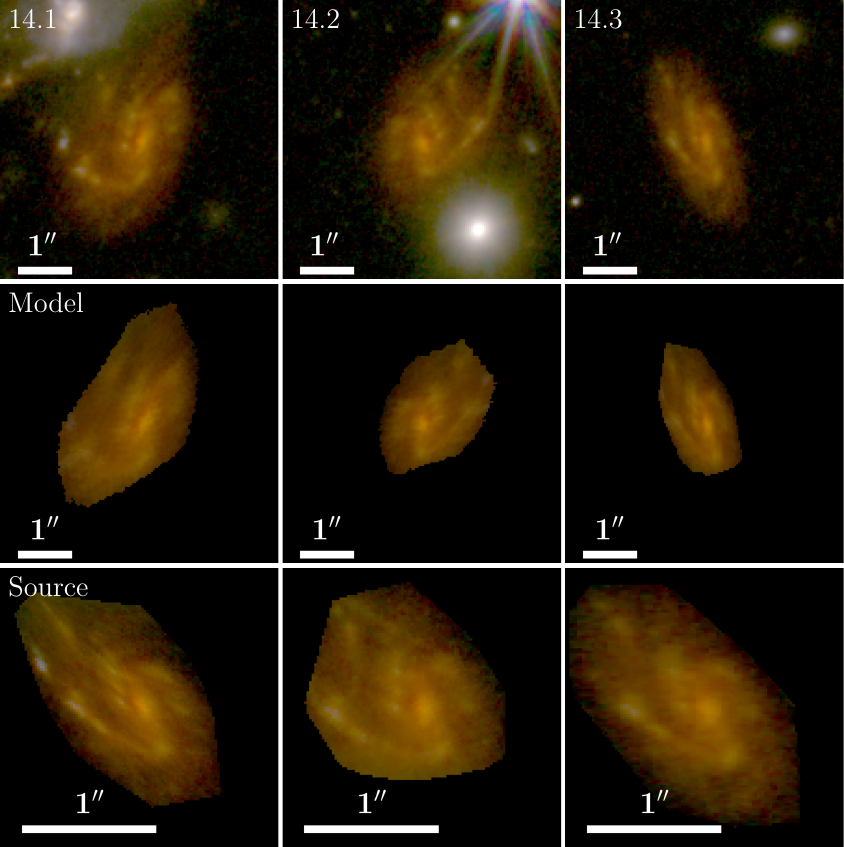}
    \caption{Reconstruction of System 14 Charybdis, the triply imaged spiral galaxy at redshift $z \simeq 3.5$, from our lens model. The first row presents the extracted observations from Fig.\,\ref{fig:Colour_image_m1931}; the second row the image reconstruction from the averaged source, for a redshift $z_{\rm 14, model} = 3.65_{-0.09}^{+0.04}$; the third row presents the source plane reconstruction for each individual image. The three images 14.1, 14.2 and 14.3 (first row, of respective magnifications $5.5 \pm 1.1$, $2.9 \pm 0.6$ and $2.4 \pm 0.5$) are projected in the source plane (third row) and averaged (see Fig.\,\ref{fig:source_reconstruction_Sys14}), and then reprojected in the lens plane (second row). The reproductions demonstrate the quality of the lens model.}
    \label{fig:sys14}
\end{figure}

\begin{figure}
    \centering
    \includegraphics[width=\linewidth]{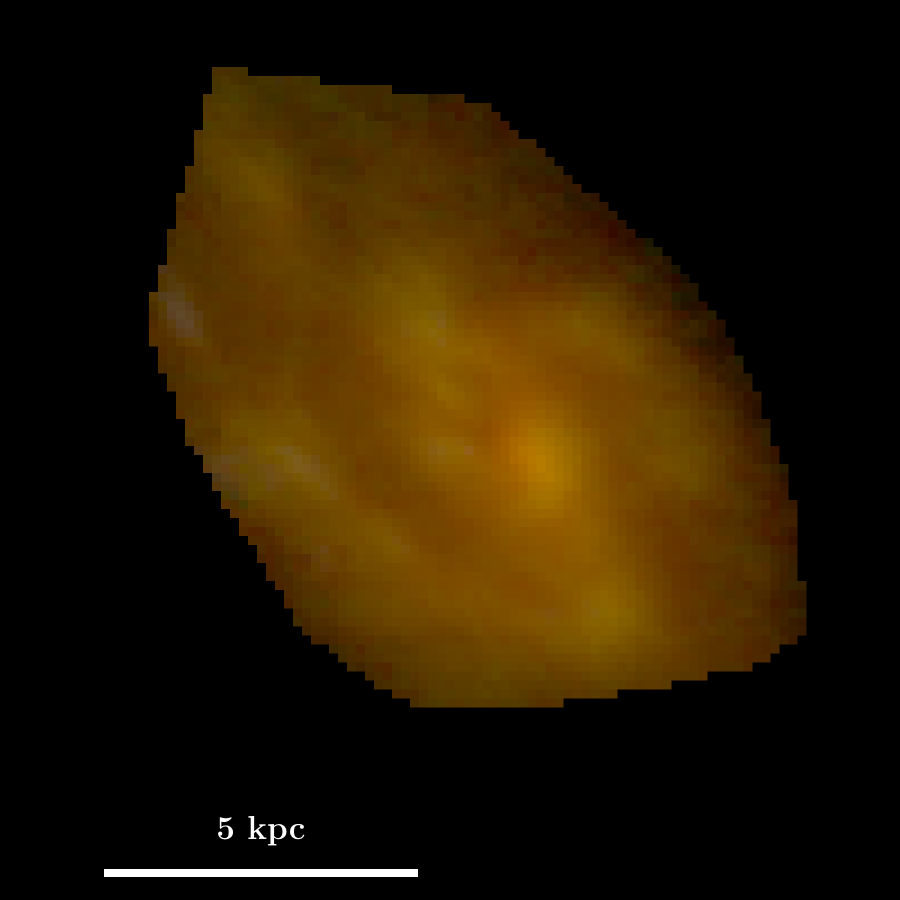} 
    \caption{Averaged source reconstruction of the spiral galaxy Charybdis at redshift $z_{\rm 14, model} = 3.65_{-0.09}^{+0.04}$, i.e.\ the triply-lensed System 14. This image was obtained by centering and averaging the source reconstruction from each individual image (and is thus effectively smoothed; cf. third row in Fig.\,\ref{fig:sys14}). The source WCS coordinate position is at $(292.9579, 26.5768)$ J2000 degrees. A physical scale bar is overplotted, where at the presumed redshift of the source of $z=3.65$,  $1'' \simeq 7.2$\,kpc.}
    \label{fig:source_reconstruction_Sys14}
\end{figure}

To further exhibit the robustness of the lens model, we show the reconstruction of a few multiple-image systems. Figure\,\ref{fig:mulim1_mulim2}, for example, shows the reconstruction of Systems 1 and 2 by the model. 
First, an image is sent to the source plane using the best-fit model. Then, this source is relensed to reproduce the images of the system, showing great consistency with the observed images. 

Perhaps more interesting is the reproduction of System 14 in Fig.\,\ref{fig:sys14} (see also Fig.\,\ref{fig:Colour_image_m1931} and Table\,\ref{tab:m1931_mulim}). 
This system was first identified as M1931-ID47/55/61 through a combination of ALMA, \textit{Herschel} and IRAC infra-red imaging \citep{Sun2022ApJ...932...77S, 2024ApJS..275...36F}, with far infra-red photometry (FIR) yielding $2 \lesssim z_{\rm phot} \lesssim 4$.
We further investigate it under the lens of JWST VENUS observations as a triply imaged\footnote{The existence of a fourth image, immediately north to the BCG but `hidden' by the intra-cluster light, remains ambiguous.} spiral galaxy, at a redshift $z_{\rm phot} \simeq z_{\rm model} \simeq 3.3 - 3.7$, in line with previous FIR photometry.\footnote{We note that all values quoted hereafter in this section are taking the redshift uncertainty into account.} We nickname the galaxy \textit{Charybdis}, after the marine monster from the Greek mythology who was believed to form whirlpools in the sea.
We present an averaged source reconstruction of the spiral galaxy in Fig.\,\ref{fig:source_reconstruction_Sys14}, generated by projecting each of the three images into the source plane (bottom row, Fig.\,\ref{fig:sys14}), aligning them and averaging the frames.\footnote{Note that in principle, the averaging of the three sources smooths the data compared to e.g.\ the most highly magnified image, but is useful for the image reproduction seen in Fig.\,\ref{fig:sys14}.}
The following re-projection of the averaged source into the image plane yields the second row of Fig.\,\ref{fig:sys14}.
This allows us to estimate the (de-lensed) projected source diameter to be $\sim 2''\times 1''$ (i.e.\ $\sim 15 \times 7$\,kpc, resp.\ semi-major and semi-minor axes, at a redshift $3.65_{-0.09}^{+0.04}$).
`Grand-design' spiral galaxies, i.e.\ galaxies exhibiting clear spiral arms, and their evolution in redshift, have long drawn curiosity, as these can shed light on when and how spiral galaxies evolved \citep[e.g.][]{Law2012SpiralGal}. Galaxy Charybdis is reminiscent of similar recent high-$z$ spiral-galaxy detections now possible with JWST, such as those described in \cite{Costantin2023SpiralGal,Kuhn2024ApJ...968L..15K,Wang2025BigWheel,Jain2025spiralgal,Xiao2025SpiralGal}. Moreover, thanks to lensing magnification ($5.5\pm1.1$), the source reconstruction presented in Fig.\,\ref{fig:source_reconstruction_Sys14} allows perhaps the most detailed view of a spiral galaxy at such a redshift ($z\sim3.5-4$) to date. 
Using the typical PSF FWHM for JWST of $0.03''$, we obtain a resolution of $\simeq 90$\,pc at the source plane. 

Provided the lack of spectroscopic identification, we further explore combined NIRCam and HST/ACS photometric analysis of this galaxy.
We derive the stellar properties of galaxy Charybdis with \texttt{Bagpipes}, following the procedure detailed in Section \ref{sec:spec-modelling}. 
Provided the observed diameter of galaxy Charybdis, we use the photometric extraction within a $1.4''$ radius aperture.
The \texttt{Bagpipes} photometric redshift is $z_{\rm phot, Charybdis} = 3.4 \pm 0.1$, compatible with the \texttt{eazy} and lens model values. 
After demagnifying the flux, we find -- on average from the fits to the three demagnified images -- a galaxy stellar mass $\log_{10} (M_{\star}/\Msol) = 10.64 \pm 0.24$, a star formation rate (SFR) of $(48.5 \pm 11.5) \Msol$\,yr$^{-1}$ and a stellar mass-weighted age of $(560\pm150)$\,Myr. 
Noticeably, we do not detect the galaxy in the VLT/MUSE data, neither in continuum or emission lines, despite at least images 14.1 and 14.2 lying well within the field-of-view and 14.3 being at the edge. Assuming $z_{14} \in [3.3, 3.7]$ and that Ly-\textalpha\ escapes from the galaxy, we would expect to detect it in the range $[522, 571]$\,nm, but do not detect any significant feature there.
%
Further analysis of this intriguing lensed galaxy and its apparent spiral structure is left to future work.

\section{Discussion} \label{sec:discussion}

\subsection{Comparison with previous lens models}

We compare here the lens modeling results with previous findings. We start by comparing to the two models presented in \cite{Zitrin2015}: a parametric model (\texttt{PIEMDeNFW}), and a Light-Traces-Mass model (\texttt{LTM}) -- in which both the galaxies and dark matter follow the cluster galaxy distribution. Both models were coupled to the same grid, which was tied to the CLASH images of the field. These two models yielded respectively, $r.m.s. = 0.77''$ and $2.28''$ (the \texttt{LTM} model has great prediction power but can be less flexible, often leading to a higher $r.m.s.$). We also compare to \cite{Caminha2019}, where seven different parametric models were presented, constructed using \texttt{Lenstool}, each adopting distinct physical assumptions. Their best $r.m.s.$ model, of $0.34''$, included two DMHs. Nevertheless we find no convincing physical prior to motivate two DMHs. Our model, including one single DMH model and a free BCG, has a $r.m.s. = 0.44''$ despite adopting fewer degrees of freedom and accommodating 12 more constraining multiple-image systems.
Our model suggests an Einstein radius, for a source at $z_s = 2$,  of $\theta_{E, z_s = 2} = 22.4 \pm 2.2 ''$. This compares very well with the \texttt{LTM} and \texttt{PIEMDeNFW} values of $22.7\pm 2.3''$ and $21.8\pm 2.2''$, thus confirming previous estimates. Similarly, the enclosed mass within these models' Einstein radii, $M_{\rm 2D} (\in \mathcal{A}_{E, z_s = 2})$ of ($8.3\pm1.3$) and ($7.8\pm1.2) \times 10^{13} \Msol$ (resp.\ for the \texttt{LTM} and \texttt{PIEMDeNFW} models), are within the $1\sigma$ estimate from our model, $(8.0 \pm 1.2) \times 10^{13} \Msol$.
We further compare our model with that of \cite{Caminha2019} through the projected mass within $200$\,kpc, of $M_{\rm 2D} (R < 200\,\mathrm{kpc}) = (1.70 \pm 0.02) \times 10^{14} \Msol$, which lies within the $1\sigma$ error range of our model $(1.64 \pm 0.25) \times 10^{14} \Msol$.
Our lens model thus confirms and further refines previous lens models for the cluster.

\subsection{Strong lensing and distant supernovae}

Cluster strong lensing offers at least two major advantages for detecting and studying high-redshift SNe. First, lensing magnification allows us to see farther and fainter, so that higher-redshift sources can be detected. Second, image multiplicity and respective time delays offer several (delayed) appearances of the same background supernova and thus greater chances for detecting it. 
In the instance of \snname, the combined magnification on both observed images increases the received flux by a factor $\mu_{101.1} + \mu_{101.2} \simeq 55$, thus decreasing the magnitude by 4.4 mag in total (2.75 per image, both being measured at $25.26-25.27$ AB mag e.g.\ in VENUS NIRCam/F410M image). This renders the object clearly observable, where without lensing, it would have been much fainter in NIRCam, and difficult to detect with NIRSpec.

Moreover, lensing time delays provide a unique outlook on transients, such that in cases where the time delays are small enough (for example, lensing by a galaxy or galaxy in a cluster), the light curve is sampled in various points in time with each single observation. This facilitates constraints on the SN type and, as was already shown in at least one case \citep[][]{Chen2022Natur.611..256C}, on the early light curve evolution and thus type and radius of the pre-explosion star.

Indeed, as was also previously illustrated on e.g.\ SNe Refsdal \citep{Kelly2015Sci,Rodney2016_SN_Refsdal}, Encore \citep{Pierel2024SN_Encore} or H0pe \citep{Frye2024SN_H0pe}, the time delays allow to both confirm the transient nature of the object and to compare different stages of the evolution. 
In the particular case of \snname, possibly for the first time, the transient nature of the object was discovered primarily not by a difference image comparing the field with previous observations, but due to the two images observed simultaneously, together with the absence of the other predicted images.

Given the current rate of multiply imaged SNe detected behind galaxy clusters with JWST (order of a handful of SNe in the few-dozen visits on galaxy clusters taken to date with JWST), we can expect many more cluster-lensed supernovae in the near future. This is in addition to yields from upcoming wide-field instruments such as the Vera Rubin Observatory and the \textit{Roman Space Telescope} \citep[see forecasts in e.g.,][]{OguriMarshall2010,Wojtak2019SN,Bronikowski2025}.

\begingroup
\renewcommand{\arraystretch}{1.2}
\begin{table*}
\centering
\hspace{-2cm}
    \begin{tabular}{lccccccc}
    \hline
    \hline
    Image & $\alpha$ & $\delta$ & $\left|\mu\right|$ & $\tau$\\
     & [deg, J2000] & [deg, J2000] & & [years, observer] \\
    \hline
    101.1 & 292.9550834 & -26.5748399 & $25.79 \pm 5.31$ & $0$ \\
    101.2 & 292.9549276 & -26.5746208 & $27.49 \pm 5.67$ & $-0.37_{-5.00}^{+0.37}$\,days \\
    p101.3 & 292.9516366 & -26.5780971 & $14.52 \pm 2.92$ & $-3.63 \pm 0.73$ \\
    p101.4 & 292.9534086 & -26.5793231 & $11.90 \pm 2.39$ & $-3.41 \pm 0.68$ \\
    p101.5 & 292.9683577 & -26.5759972 & $2.16 \pm 0.43$ & $-53.94 \pm 10.79$ \\
    \hline
    \hline
    \end{tabular}
    \caption{Summary of detected and predicted images of \snname. Predicted images are preceded with a `p'. Columns are: (i) Object Id; (ii) \& (iii) R.A.\ and Dec.\ in degrees; 
    (iv) Lensing magnification $\mu$ for the best-performing model; (v) Relative time delay $\tau$ in the observer frame, measured from the observation of image 101.1, in years. Image 101.1 was the last one to arrive. The $\mu$ and $\tau$ values were computed at the predicted position, and we provide the average values of the sample of well-converged models. We take the systematic errors to be $\sim 20\%$ for both $|\mu|$ and $\tau$. 
    The model variability on the time delay between the observed images 101.1 and 101.2 is larger, however, and we infer from a few models with different physical priors a conservative $5$ days systematic, respective of the order of image arrival (101.2 must arrive before 101.1 due to topological considerations). 
    We here present the cumulated uncertainties, where the systematic dominate statical errors.}
    \label{tab:TD_SNEos}
\end{table*}
\endgroup

\subsection{Cosmological constraints from lensed supernovae}

The spectroscopic detection of \snname at $z=5.133$, i.e.\ the most distant multiply-lensed supernova known to date, motivates a discussion on the possible use of lensed supernovae (or other transient) observations to constrain cosmology, and, in particular, whether high-redshift lensed supernovae are more preferable for that purpose.

\subsubsection{The Hubble constant} \label{sec:H0_TD_opt}

One of the primary methods for constraining the expansion rate of the Universe is through the analysis of lensing time delays. The measurement of the time delays between images of a source, when coupled with a lens model, provides a direct constraint on the Hubble constant $H_0$ through: 
\begin{equation}
    \Delta t = \frac{1 + z_l}{c} \frac{D_l D_s}{D_{ls}} \left[ \tau(\boldsymbol{\theta}_1, \boldsymbol{\beta}) - \tau(\boldsymbol{\theta}_2, \boldsymbol{\beta}) \right],
    \label{eq:time-delays}
\end{equation}
where $\Delta t$ is the measured, observed time delay between two images (indices 1 and 2), $z_l$ the lens redshift, $D_l$, $D_s$ and $D_{ls}$, respectively, the angular distance of the lens, of the source, and between the lens and source. The Fermat potential for an image of position $\boldsymbol{\theta}_i$ of a source at position $\boldsymbol{\beta}$ is derived from the lens model and expressed as:
\begin{equation}
    \tau (\boldsymbol{\theta}, \boldsymbol{\beta}) = \frac{(\boldsymbol{\theta} - \boldsymbol{\beta})^2}{2} - \Psi (\boldsymbol{\theta}),
    \label{eq:Fermat_potential}
\end{equation}
where the first term represents the geometric time delay and the second, $\Psi$, the Shapiro, or gravitational delay.
We note that the Hubble constant scales as:
\begin{equation}
    H_0 \propto D_{\Delta t}^{-1}, 
    \label{eq:H0_TD_distance}
\end{equation}
where $D_{\Delta t}$ represents the time-delay distance \citep[cf.\ e.g.][]{Refsdal1964MNRAS, Birrer2024TD}:
\begin{equation}
    D_{\Delta t} = (1 + z_l) \frac{D_l D_s}{D_{ls}}.
    \label{eq:time-delay_distance}
\end{equation}
Since the lens model does not depend on the Hubble constant, simply comparing the measured time delays with the difference in the Fermat potential at the image positions supplies a direct constraint on $H_0$. 

We wish to examine whether the typical time delay increases or decreases with source redshift. 
Using the best-model of cluster \clname as a simulated cluster, we project randomly drawn background sources at redshifts $z_s \in \left\{ 2, 5, 8\right\}$ into the image plane and measure for each system the time delays $\Delta t$ between its multiple images.
We find the typical time delays to grow, on average, with source redshift, thus confirming results from other studies \citep[see][and references therein]{Golubchik2024ApJ...976..108G}.
The relative time delay should thus be measurable with greater precision for high-redshift sources, placing tighter constraints on the Hubble constant.

\subsubsection{Other cosmological parameters} \label{subsec:cosmography_TD_constraints}

The time-delay distance, $D_{\Delta t}$, depends not only on the Hubble parameter but also on other cosmological parameters, such as matter density $\Omega_{m}$, dark energy density $\Omega_{\Lambda}$, and its equation of state parameters $w_0$, $w_a$, in e.g.\ the CPL parametrization by \citet{ChevallierPolarski2001DarkEnergy,Linder2003PhRvL..90i1301L}. 
This means that time delays can also be used, in principle, to put constraints on these parameters. But, in contrast to $H_0$, the lens model \textit{does} depend on these parameters. Thus, the correct way to use measured time delays to constrain cosmological models is to leave these parameters free in the lens modeling, while using the measured time delays as constraints \citep[as e.g.][]{Caminha2022}. 
We do not undertake such an analysis here, especially given that the time delay between both detected images is very short ($<5$ days, see Table \ref{tab:TD_SNEos}) and consistent with zero. 
Nevertheless, provided the high redshift of \snname, it remains compelling to examine whether high-redshift, multiply imaged supernovae could offer advantages for cosmography.

To probe this, we run two additional lens models for \clname, with a setup identical to that presented in Section\,\ref{sec:analysis}, but with substantially different (fixed) cosmological models: in one model, we fix $\Omega_m = \Omega_{\Lambda} = 0.5$ (all other cosmological parameters being equal); and in the second model, we fix the linear dark energy equation of state parameter to be $w_0 = -0.5$ (as opposed to $w_0 = -1$ in \LCDM). We keep $w_a = 0$ in all. The resulting models converged with convincing precision (resp.\ $r.m.s. = 0.51''$ and $0.57''$), and the time delays derived for \snname in both scenarios are -- unsurprisingly, due to the small delays -- also compatible with the measurements within the uncertainties. We randomly simulate 10,000 sources at redshifts $z_s \in \left\{ 2, 5, 8 \right\}$ for each of the three models, reproject them onto the lens plane and measure the respective time delays between the multiple images of each. 
We find that larger discrepancies in the median time delay between the three models are obtained for lower-redshift sources (close to $\sim$10\%). 
Nevertheless, even higher-redshift sources -- such as those at a redshift similar to \snname\ -- retain substantial discriminating power (about $\sim5\%$).

We also examine the expected change in the time-delay distance analytically and verify that indeed $\sim5-10\%$ variations are expected between the three probed models. The cosmological model variations included in the time-delay distance $D_{\Delta t}$ are vastly dominated by the variations in $D_l$ ($\sim 7\%$ difference between the three cosmologies for the redshift of \clname), and variations in $D_s/D_{ls}$ across the three cosmological models, for $z_s > 2$, are more limited, contributing $\lesssim 3\%$. We conclude that the time-delay distance can be used to constrain cosmology, but the gain from a higher-redshift supernovae may not be substantial in this case. 
This is unlike the $H_{0}$ case for which higher-redshift supernovae may be preferable. 
These preliminary conclusions are based on a few distinct cases. A more thorough investigation, including probing cosmological parameters such as $\Omega_{m}$ and $w_{0}$ using measured time delays at different source redshifts, and the susceptibility to various systematics (lens model, number of images, etc.) is left to future studies.

\section{Summary} \label{sec:summary}

In this work we construct a new lens model for the galaxy cluster \clname, accompanying the detection of \snname at $z=5.133$, using a combination of newly acquired JWST/NIRCam observations from the VENUS survey, ancillary HST imaging from the CLASH survey, and spectroscopic data from MUSE and NIRSpec. 
We stress that \snname was first detected through the cluster lens model: the two red point-like sources, observed in VENUS/NIRCam data next to the BCG, were designated as two likely images of a $z\sim5$ supernova, due to the short time delay between them and lack of other images where the model predicts.
The model is finally constrained using 19 multiply-lensed sources, including 10 newly identified systems and a total of 9 with a spectroscopic redshift, and has a very good image reproduction $r.m.s.$ of \rmsfinal.

Motivated by the detection of \snname at $z=5.13$, we discuss whether high-redshift SNe are also useful for constraining $H_{0}$ and other cosmological parameters through the time-delay distance between the images of a multiply imaged source. 
We simulate how, as time delays increase on average for higher-redshift sources, a more precise $H_{0}$ measurement could be obtained. In contrast, we find that, at least for the few particular cases we ponder here, for other cosmological parameters (e.g.\ the matter density or equation of state), lower-redshift sources provide moderately stronger cosmographic constraints, yet that higher-redshift sources similar to \snname can still supply sufficient constraining power. 
Regular and frequent monitoring of strong-lensing galaxy clusters such as \clname is expected to yield substantial samples of multiply-imaged transient sources over the next decade, enabling cosmography from measured time delays and distance ratios in strong lensing clusters.

One of the systems we find behind \clname is a triply-lensed grand-design spiral galaxy, which we name \textit{Charybdis}, at redshift $z \simeq 3.5$, observed with exceptional resolution due to lensing. Indeed, with a magnification of $\simeq5.5\pm1.1$ for the brightest image, we can reach a $\lesssim 100$\,pc resolution in the source plane. We produce a detailed source image and examine its physical properties using \texttt{Bagpipes}, deriving a high stellar mass ($4.4_{-1.9}^{+3.2} \times 10^{10} \Msol$) and SFR ($48.5 \pm 11.5 \,\Msol$\,yr$^{-1}$).

These results add to the already impressive display of the VENUS science output, spanning from lensed high-redshift galaxies \citep{Nakane2025_VENUS_highzgal, Abdurrouf2025}, LRDs \citep{Golubchik2025_LRD_VENUS, Zhang2025LRD, Yanagisawa2026LRD}  and of course supernovae (such as \snname, cf.\ \citealt{Coulter2026_SN_Eos}) and Ly-\textalpha\ emitting galaxies at high-redshift (Eos host galaxy, cf.\ Y.\ Asada et al.\ in prep.). These findings foreshadow once more the impressive legacy of JWST for the study of gravitational lensing, cosmological transients and the high-redshift Universe.

\begin{acknowledgments} 
The authors acknowledge the use of the Canadian Advanced Network for Astronomy Research (CANFAR) Science Platform operated by the Canadian Astronomy Data Center (CADC) and the Digital Research Alliance of Canada (DRAC), with support from the National Research Council of Canada (NRC), the Canadian Space Agency (CSA), CANARIE, and the Canadian Foundation for Innovation (CFI).
This work is based on observations made with the NASA/ESA/CSA James Webb Space Telescope and with the NASA/ESA Hubble Space Telescope. The data were obtained from the \texttt{Mikulski Archive for Space Telescopes} (\texttt{MAST}) at the \textit{Space Telescope Science Institute} (STScI), which is operated by the Association of Universities for Research in Astronomy, Inc., under NASA contract NAS 5-03127 for JWST. These observations are associated with JWST programs \#6882 (VENUS) and 9493 (\snname) and HST \#12456 (CLASH).
The HST specific observations analyzed can be accessed via \dataset[doi: 10.17909/vzxq-a559]{https://doi.org/10.17909/vzxq-a559}, while the JWST data set is at \dataset[doi: 10.17909/swh2-9h05]{https://doi.org/10.17909/swh2-9h05}.
The maps associated to the lens model, as well catalogs cited in this article may be downloaded from the Zenodo dataset \dataset[doi: 10.5281/zenodo.18646525]{https://doi.org/10.5281/zenodo.18646525}.

AZ acknowledges support by the Israel Science Foundation Grant No.\ 864/23. 
LJF acknowledges support from the University of Texas at Austin Cosmic Frontier Center.
RAW acknowledges support from NASA JWST Interdisciplinary Scientist grants NAG5-12460, NNX14AN10G and 80NSSC18K0200 from GSFC.
PD warmly acknowledges support from an NSERC discovery grant (RGPIN-2025-06182).
PL thanks the support from CNPq, grants 310260/2025-6 and 404160/2025-5, and FAPERJ, grant E-26/200.545/2023.
GEM acknowledges the Villum Fonden research grants 37440 and 13160. The Cosmic Dawn Center (DAWN) is funded by the Danish National Research Foundation under grant DNRF140.
MB, NM and GR acknowledge support from the ERC Grant FIRSTLIGHT \#101053208 and Slovenian national research agency ARIS through grants N1-0238 and P1-0188.
FEB acknowledges support from ANID-Chile BASAL CATA FB210003 and FONDECYT Regular 1241005.
YJ-T acknowledges financial support from the State Agency for Research of the Spanish MCIU through Center of Excellence Severo Ochoa award to the Instituto de Astrof\'isica de Andaluc\'ia CEX2021-001131-S funded by MCIN/AEI/10.13039/501100011033, and from the grant PID2022-136598NB-C32 Estallidos and project ref. AST22-00001-Subp-15 funded by the EU-NextGenerationEU.

This work made use of the following software packages: \texttt{Astropy} \citep{astropy:2013}, \texttt{Matplotlib} \citep{Hunter:2007}, \texttt{NumPy} \citep{numpy}, \texttt{pandas} \citep{mckinney-proc-scipy-2010}, \texttt{Python} \citep{python}, \texttt{SciPy} \citep{2020SciPy-NMeth}, \texttt{Cython} \citep{cython:2011}, \texttt{emcee} \citep{emcee-Foreman-Mackey-2013,emcee_10996751}, \texttt{Numba} \citep{numba:2015,Numba_17874185}, and \texttt{h5py} \citep{collette_python_hdf5_2014}.  

\end{acknowledgments}


\newpage
\bibliographystyle{aasjournalv7}
\bibliography{MyBiblio}

\appendix

\section{Multiple images table}

\startlongtable
\tabletypesize{\footnotesize}
\begin{deluxetable*}{lccccc}
\tablecaption{\small{Cluster \clname: Multiple Images and Candidates}}
\label{tab:m1931_mulim}
\tablehead{
\colhead{Id.} &
\colhead{$\alpha$} &
\colhead{$\delta$} &
\colhead{$z_{\rm s}$} &
\colhead{$z_{\rm model}$} &
\colhead{Comments}}
\startdata
\hline
\hline
1.1 & $292.9556015$ & $-26.5741522$ & $1.178$ & $1.178$ & \\ 
1.2 & $292.9561018$ & $-26.5774723$ & $1.178$ & & \\ 
1.3 & $292.9521049$ & $-26.5761848$ & $1.178$ & & \\ 
1.4 & $292.9639586$ & $-26.5751975$ & $1.178$ & & \\ 
\hline 
2.1 & $292.9579777$ & $-26.5686133$ & $1.835$ & $1.835$ & \\ 
2.2 & $292.9605911$ & $-26.5691827$ & $1.835$ & & \\ 
2.3 & $292.9496095$ & $-26.5706026$ & $1.835$ & & \\ 
\hline 
3.1 & $292.9519651$ & $-26.5827949$ & $2.707$ & $2.707$ & \\ 
3.2 & $292.9554405$ & $-26.5836838$ & $2.707$ & & \\ 
3.3 & $292.9646857$ & $-26.5817857$ & $2.707$ & & \\ 
\hline 
4.1 & $292.9529960$ & $-26.5835709$ & $4.000$ & $4.000$ & \\ 
4.2 & $292.9539349$ & $-26.5838012$ & $4.000$ & & \\ 
4.3 & $292.9655055$ & $-26.5819623$ & $4.000$ & & \\ 
\hline 
5.1 & $292.9531783$ & $-26.5713373$ & $4.745$ & $4.745$ & \\ 
5.2 & $292.9504752$ & $-26.5732984$ & $4.745$ & & \\ 
\hline 
6.1 & $292.9607683$ & $-26.5786357$ & $5.079$ & $5.079$ & \\ 
6.2 & $292.9621377$ & $-26.5769782$ & $5.079$ & & \\ 
6.3 & $292.9583435$ & $-26.5740771$ & $5.079$ & & \\ 
\hline 
7.1 & $292.9527960$ & $-26.5728090$ & $5.339$ & $5.339$ & \\ 
7.2 & $292.9511559$ & $-26.5744529$ & $5.339$ & & \\ 
\hline
8.1 & 292.9550587 & -26.5679326 & $\left[2.34^{+0.15}_{-2.14}\right]$ & $2.45_{-0.03}^{+0.03}$ &  \\
8.2 & 292.9512212 & -26.5690951 & $\left[2.23^{+0.06}_{-0.18}\right]$ & &  \\
8.3 & 292.9622701 & -26.5688525 & $\left[2.12^{+0.21}_{-0.30}\right]$ & &   \\
\hline
9.1 & 292.9555356 & -26.586321 & $\left[1.92^{+2.54}_{-0.52}\right]$ & $19.03_{-4.86}^{+0.57}$ & $z_{\rm model}$ \\
9.2 & 292.9537652 & -26.5860859 & $\left[3.07^{+0.12}_{-0.09}\right]$ &  & poorly constrained \\
\hline
10.1 & 292.9547204 & -26.5728671 & $\left[0.43^{+0.09}_{-0.06}\right]$ & $0.66_{-0.00}^{+0.00}$ &   \\
10.2 & 292.9568584 & -26.5726138 & $\left[0.54^{+0.09}_{-0.02}\right]$ &  &   \\
\hline
11.1 & 292.9525117 & -26.5682155 & \_ & $2.28_{-0.06}^{+2.28}$ & Photometric analysis \\
11.2 & 292.954243 & -26.5677604 & \_ &  &  does not constrain \\
c11.3 & 292.9585968 & -26.5683021 & \_ &  & the redshift  \\
\hline
12.1 & 292.9575243 & -26.5862062 & $\left[3.03^{+0.06}_{-0.02}\right]$  & $1.97_{-0.08}^{+0.01}$ &   \\
12.2 & 292.9559794 & -26.5863523 & $\left[2.49^{+1.89}_{-1.16}\right]$ &  &   \\
12.3 & 292.9611313 & -26.5859852 & $\left[2.45^{+1.24}_{-1.56}\right]$ &  &   \\
\hline
13.1 & 292.9586182 & -26.5890125 & \_ & $4.05_{-0.44}^{+0.14}$ &   \\
13.2 & 292.9571188 & -26.5890165 & \_ &  &   \\
13.3 & 292.9578123 & -26.5890353 & \_ & &   \\
\hline
14.1 & 292.9503238 & -26.5787239 & $\left[3.43^{+0.07}_{-0.13}\right]$ & $3.65_{-0.09}^{+0.04}$ &  Lensed grand-design spiral\\
14.2 & 292.955475 & -26.580854 & $\left[3.41^{+0.06}_{-0.41}\right]$ & & \textbf{galaxy Charybdis} \\
14.3 & 292.9669864 & -26.5767691 & $\left[3.42^{+9.39}_{-0.03}\right]$ &  &  See Fig.\,\ref{fig:sys14}, \ref{fig:source_reconstruction_Sys14} \\
\hline
15.1 & 292.9610451 & -26.5766429 & $\left[5.87^{+0.30}_{-0.27}\right]$ & $5.37_{-1.12}^{+1.84}$ &   \\
15.2 & 292.9608161 & -26.5770511 & $\left[6.44^{+0.15}_{-0.29}\right]$ &  &   \\
\hline
16.1 & 292.9542332 & -26.5824673 & $\left[7.55^{+0.12}_{-0.25}\right]$ & $7.82_{-0.36}^{+0.19}$ & \\
16.2 & 292.9510206 & -26.5811211 & $\left[7.48^{+0.40}_{-0.38}\right]$ &  &   \\
16.3 & 292.9676024 & -26.5786247 & $\left[7.44^{+0.26}_{-0.30}\right]$ &  &   \\
\hline
17.1 & 292.9523087 & -26.5732621 & $\left[4.73^{+3.65}_{-2.81}\right]$ & $8.43_{-0.98}^{+1.90}$ &  $z_{\rm phot}$ poorly constrained \\
17.2 & 292.9515177 & -26.574158 & $\left[4.64^{+2.02}_{-3.61}\right]$ &  &   \\
\hline
c18.1 & 292.9628796 & -26.5684558 & $\left[2.09^{+0.18}_{-0.19}\right]$ & $10.02^{+4.42}_{-2.46}$ & Not used as constraints; \\
c18.2 & 292.9475661 & -26.5713572 & $\left[1.96^{+0.13}_{-0.15}\right]$ &  & $z_{\rm model}$ derived \textit{a posteriori}\\
\hline
100.1 & 292.9551274 & -26.5748739 & 5.133 & 5.133 & Host \textbf{Eos galaxy}   \\
100.2 & 292.9548835 & -26.574538 & 5.133 & & (see Y.\ Asada et al., in prep.) \\
100.3 & 292.9516473 & -26.5782475 & 5.133 & &   \\
100.4 & 292.9538717 & -26.579633 & 5.133 & &   \\
\hline
101.1 & 292.9550834 & -26.5748399 & 5.133 & 5.133 & \textbf{\snname}   \\
101.2 & 292.9549276 & -26.5746208 & 5.133 & & \citep[see][]{Coulter2026_SN_Eos}  \\
\hline
\hline
\enddata
\tablecomments{Columns: (i) Arc Id. `c' stands for candidate where identification was more ambiguous, or if image was not used as constraint. Systems 1-7 were already included in \cite{Caminha2019}, and systems 2, 3, 4, 8 and 9 were identified in \citet{Zitrin2015}. Systems 10 to 18, as well as the Eos systems (100, 101) are newly identified.
(ii) and (iii) R.A.\ and Dec.\ in degrees, J2000.
(iv) Source redshift. Systems 1-7, 100 and 101 have spectroscopic redshifts, given without uncertainties. \texttt{eazy} photometric redshifts and their $1\sigma$ uncertainties, for systems lacking a spectroscopic measurement, are designated with brackets. 
(v) Redshift and $1\sigma$ uncertainties from the lens model.
(vi) Comments.
System 18 is a candidate, and thus not used as a constraint. We still provide its redshift from the best-fit model. Note system Eos has other predicted images (see Table \ref{tab:TD_SNEos}).
}
\end{deluxetable*}

\section{Density profiles in \softname} \label{sec:pot_in_astrolemonpy}

\subsection{Expression of the ellipticity}

In \softname, the 2D mass ellipticity is parametrized using the flattening $f = 1 - r_b/r_a = 1 -q$, for a semi-major (minor) axis $r_a$ ($r_b$), and the position angle $\theta$. They are incorporated into the elliptical radius $R$. For a point at circular radius $R = (x, y)$, the elliptical coordinate transformation is:
\begin{equation}
    R^2 =  \left[ x \cos \theta + y \sin \theta\right]^2 - q^{-2} \left[ x \sin \theta - y \cos \theta\right]^2.
\end{equation}

\subsection{Density profiles}

We present here the expressions for the density profiles implemented in \softname, and presented in Section \ref{sec:analysis}.
For dark-matter halos (DMH), we generally prefer the un-truncated Pseudo-Isothermal Elliptical Matter Distribution (PIEMD, \citealt{Keeton2001models}). We give its 3D mass density: 
\begin{equation*}
    \rho (r) = \frac{\rho_0}{1 + \left(r / a \right)^2} = \frac{\sigma_0^2}{2 \pi G} \frac{1}{a^2 + r^2},
    \label{eq:PIEMD_profile}
\end{equation*}
where:
\begin{equation}
    \rho_0 = \frac{\sigma_0^2}{2 \pi G a^2},
\end{equation}
is the 3D-density normalization, $a$ the core radius, $\sigma_0$ the 1D velocity dispersion and $G$ the constant of Newton. In the thin lens approximation, we only constrain the projected density:
\begin{equation}
    \Sigma (R) = \Sigma_0 \frac{a}{\sqrt{a^2 + R^2}} = \frac{\sigma_0^2}{2 G} \frac{1}{\sqrt{a^2 + R^2}},
    \label{eq:PIEMD_2D_profile}
\end{equation}
where:
\begin{equation}
    \Sigma_0 = \frac{\sigma_0^2}{2 G a} = \pi \rho_0 a,
\end{equation}
is the surface density normalization.

As for individual galaxies, we model them as a collection of dual Pseudo-Isothermal Elliptical profiles \citep[dPIE, or pseudo-Jaffe; see][]{KassiolaKovner1993, Eliasdottir2007arXiv0710.5636E}, i.e.\ a cut-off PIEMD profile: $\Theta_{\rm dPIE} (R; a, s) = \Theta_{\rm PIEMD} (R; a) - \Theta_{\rm PIEMD} (R; s)$ for any observable $\Theta$.
The dPIE 3D density is given by:
\begin{equation}
    \rho (r) = \frac{\rho_0}{\left[1 + \left(r / a \right)^2\right] \left[1 + \left(r / s \right)^2\right]} = \frac{\sigma_0^2}{2 \pi G} \left[ \left(a^2 + r^2\right)^{-1} - \left(s^2 + r^2\right)^{-1} \right],
    \label{eq:dPIE_profile}
\end{equation}
where $s > a$ is the cut-radius. Let us note that the PIEMD case is the $s \rightarrow \infty$ limit. We generalize the expression of the density normalization:
\begin{equation}
    \rho_0 = \frac{\sigma_0^2}{2 \pi G} \frac{s^2 - a^2}{a^2 s^2},
\end{equation}
which we note is slightly different from the usual approximation $s-a \approx s$ \citep{Limousin2005}.
Similarly to PIEMD, the projected density writes:
\begin{equation}
    \Sigma (R) = \Sigma_0 \frac{a s}{s - a} \left[ \frac{1}{\sqrt{R^2 + a^2}} - \frac{1}{\sqrt{R^2 + s^2}} \right] = \frac{\sigma_0^2}{2 G} \left[ \left( R^2 + a^2 \right)^{-1/2} - \left( R^2 + s^2 \right)^{-1/2} \right],
    \label{eq:dPIE_2D_profile}
\end{equation}
where:
\begin{equation}
    \Sigma_0 = \frac{\sigma_0^2}{2 G} \frac{s - a}{a s} = \pi \rho_0 \frac{a s}{a+s}.
\end{equation}
The deflection angle associated to the dPIE distribution is:
\begin{equation}
    \alpha (R) = \frac{4 \pi \sigma_0^2}{c^2} \frac{D_{ls}}{D_s} \frac{1}{R} \left[ \sqrt{a^2 + R^2} - a - \sqrt{s^2 + R^2} + s  \right],
\label{eq:alpha_dPIE}
\end{equation}
where $D_{s}$ and $D_{ls}$ are respectively the angular distance of the source and between the lens and source.
The associated 2D lensing potential such that $\boldsymbol{\alpha} = \boldsymbol{\nabla} \Psi$ is:
\begin{equation}
    \Psi (R)  = 2 \pi \sigma_0^2 \left[ \sqrt{s^2 + R^2} - \sqrt{a^2 + R^2} - s \ln \left( s + \sqrt{s^2 + R^2} \right) + a \ln \left( a + \sqrt{a^2 + R^2} \right) \right].
\label{eq:Lenspot_dPIE}
\end{equation}

We then scale all cluster-member galaxies together using the Faber-Jackson relationship for red elliptical galaxies \citep{FaberJackson}, the galaxies amounting to only two free parameters: the scale cut-radius $s_{\star}$ and velocity dispersion $\sigma_{\star}$. We fix the typical core-radius $a_{\star} = 0.2$\,kpc \citep[following the typical values of elliptical galaxies and other lens modelers][]{Wallington1993, Richard2014FF, Furtak2023UNCOVER}, although \softname allows it to remain free, if desired.
Each galaxy's (index $i$) dPIE profile is then scaled:
\begin{equation}
    \begin{split}
    \sigma_{0, i} = \sigma_{0, \star} \left( \frac{L_i}{L_{\star}} \right)^{\lambda},~~~~~~~~~~
    a_i = a_{\star} \left( \frac{L_i}{L_{\star}} \right)^{\beta},~~~~~~~~~~
    s_i = s_{\star} \left( \frac{L_i}{L_{\star}} \right)^{\alpha}.
    \label{eq:Faber-Jackson}
    \end{split}
\end{equation}
We fix here the power indices to their default Faber-Jackson values $\lambda = 1/4$, $\beta = 1/2$ and $\alpha = 1/2$, although \softname also allows us to optimize them. 
Finally, we allow the luminosity of individual galaxies, $L_i$ for galaxy $i$, to be rescaled. The associated weight $w_i$ simply multiplies $L_i$ such that $L_i \rightarrow w_i L_i$.

\section{Posterior model distribution}

\begin{figure}
    \centering
    \includegraphics[width=1\linewidth]{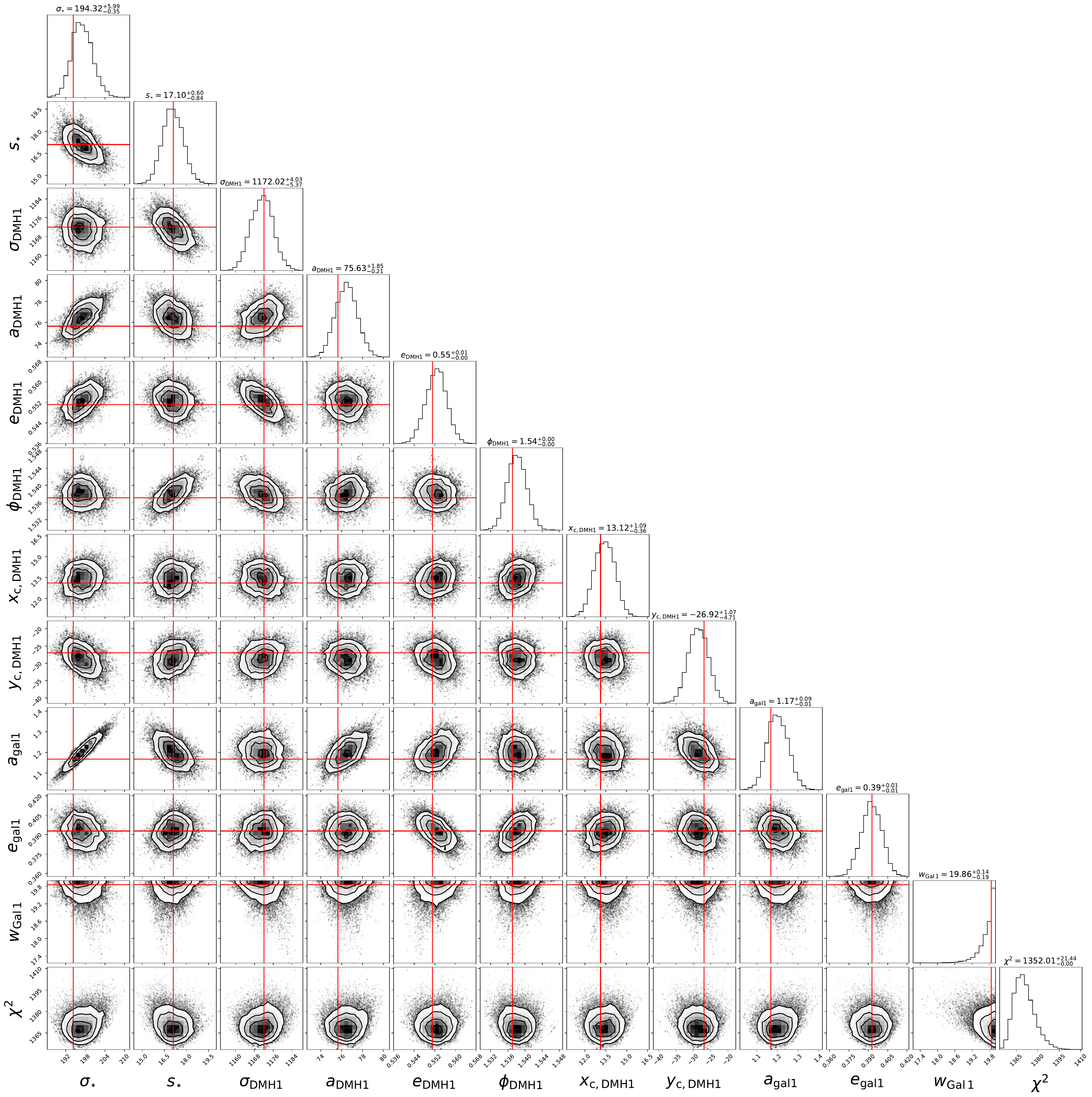}
    \caption{`Cornerplot' post-burn-in posterior distribution of the optimized parameters. 
    $e = f$ represents the flattening ellipticity. The red lines and the values above each column represent the best-fit value (see Table\,\ref{tab:best_model}).
    The BCG weight $w_{\rm Gal 1}$ is limited to 20 for physical reasons -- i.e.\ given its measured velocity dispersion. We note that when allowing for arbitrarily high values, it reaches $w_{\rm Gal 1} = 44_{-6}^{+1}$. However, given the degeneracy between the BCG and DMH, such a solution is only marginally superior (i.e., it shows a relatively similar $r.m.s\simeq0.4\arcsec$). We thus use prior knowledge of the velocity dispersion of the BCG to retain a more physical interpretation.}
    \label{fig:cornerplot_vA_52}
\end{figure}

\end{document}